\documentclass[12pt]{article}
\usepackage{epsfig, epsf, graphicx, subfigure}
\usepackage{pstricks, pst-node, psfrag}
\usepackage{amssymb,amsmath}
\usepackage{verbatim,enumerate}
\usepackage{rotating, lscape}
\usepackage{setspace}
\usepackage{multicol}
\usepackage{multirow}
\usepackage{gensymb} 
\usepackage{natbib}
\usepackage{mathtools}
\usepackage{booktabs}												
\usepackage[utf8]{inputenc}
\usepackage[colorlinks,citecolor=blue,urlcolor=blue]{hyperref}
\usepackage[]{algorithm2e}
\usepackage[toc,page]{appendix}
\usepackage{amssymb}
\usepackage{pifont}
\def\log{\hbox{log}}

\def\boxit#1{\vbox{\hrule\hbox{\vrule\kern6pt
          \vbox{\kern6pt#1\kern6pt}\kern6pt\vrule}\hrule}}

\def\bse{\begin{eqnarray*}}
\def\ese{\end{eqnarray*}}
\def\be{\begin{eqnarray}}
\def\ee{\end{eqnarray}}
\def\bq{\begin{equation}}
\def\eq{\end{equation}}
\def\bse{\begin{eqnarray*}}
\def\ese{\end{eqnarray*}}

\newcommand\independent{\protect\mathpalette{\protect\independenT}{\perp}}
\def\independenT#1#2{\mathrel{\rlap{$#1#2$}\mkern2mu{#1#2}}}

\usepackage[margin=1in,top=1in,bottom=1in]{geometry}
\allowdisplaybreaks

\begin{document}

\thispagestyle{empty} \baselineskip=28pt \vskip 5mm
\begin{center} {\Huge{\bf Bayesian Modeling of Air Pollution Extremes Using Nested Multivariate Max-Stable Processes}}
\end{center}
\baselineskip=12pt \vskip 10mm

\begin{center}\large
Sabrina Vettori$^1$,
Rapha\"el Huser$^1$,
and Marc G.~Genton$^1$
\end{center}

\footnotetext[1]{
\baselineskip=10pt King Abdullah University of Science and Technology (KAUST), Computer, Electrical and Mathematical Science and Engineering Division (CEMSE),
Thuwal 23955-6900, Saudi Arabia. \\ E-mails: sabrina.vettori@kaust.edu.sa, raphael.huser@kaust.edu.sa, marc.genton@kaust.edu.sa.}

\baselineskip=17pt \vskip 10mm \centerline{\today} \vskip 10mm


\begin{center}
{\large{\bf Abstract}}
\end{center}

\baselineskip=17pt

\noindent 


Capturing the potentially strong dependence among the peak concentrations of multiple air pollutants across a spatial region is crucial for assessing the related public health risks. In order to investigate the multivariate spatial dependence properties of air pollution extremes, we introduce a new class of multivariate max-stable processes. Our proposed model admits a hierarchical tree-based formulation, in which the data are conditionally independent given some latent nested $\alpha$-stable random factors. The hierarchical structure facilitates Bayesian inference and offers a convenient and interpretable characterization. We fit this nested multivariate max-stable model to the maxima of air pollution concentrations and temperatures recorded at a number of sites in the Los Angeles area, showing that the proposed model succeeds in capturing their complex tail dependence structure.

\par\vfill\noindent
\par\medskip\noindent

\clearpage\pagebreak\newpage \pagenumbering{arabic}
\baselineskip=26pt

\section{Introduction}\label{sec:intro}
Modeling the joint behavior of multivariate extreme events is of interest in a wide range of applications, ranging from finance and telecommunications to Earth and environmental sciences, such as hydrology, seismology, and applications related to climate change or air pollution monitoring.  
Simultaneous exposure to multiple air pollutants seriously affects public health worldwide, causing loss of life and livelihood and requiring costly health care. Therefore, policymakers such as those at the US Environmental Protection Agency (EPA), are researching multivariate approaches to quantify air pollution risks  \citep{Dominici2010,Johns2012}.
The issue of air pollution is compounded by global warming and climate change, as increasingly high temperatures are suspected to contribute to raising ozone concentrations and aggravating their effect in the human body \citep{Kahle2015}. 
This pressing situation urges a greater understanding and better monitoring of air pollution extremes, the complexity of which poses a challenge for standard statistical techniques. Indeed, multiple air pollutants are often recorded at various spatial locations and, therefore, the modeling of peak exposures across a spatial region must transcend the assumption of independence in order to capture their spatial variability. In this paper, we propose a new methodological framework based on Extreme-Value Theory, for estimating the probability that multiple air pollutants and temperatures will be simultaneously extreme at different spatial locations. 
 
The statistical modeling of single extreme variables observed over space is usually based on spatial max-stable processes (see, e.g., the reviews by \citealp{Davison2012b}, \citealp{Davison2015}, and \citealp{Davison.etal:2018}), which are the only possible limit models for properly renormalized block maxima and whose block size increases to infinity. Except in the case of asymptotic independence, these models can be used to capture the potentially strong spatial dependence that may exist among variables at extreme levels. 
Models for asymptotically independent data in the bivariate case were introduced by \citet{Ledford1996}; more recently, \citet{Wadsworth2012}, \citet{Optiz2016}, \citet{Huser2017a}, and \citet{Huser2018} proposed several spatial models for high threshold exceedances that can handle both asymptotic dependence and independence; see also the related paper by \citet{Krupskii.etal:2018}. However, the current literature on multivariate modeling of spatial extremes is still rather sparse. Here, we restrict ourselves to asymptotic dependence by modeling multivariate block maxima recorded over space using a suitable max-stable process. 
\citet{Genton2015} proposed multivariate versions of the Gaussian \citep{Smith1990}, the extremal-Gaussian \citep{Schlather2002b}, extremal-$t$ \citep{Nikoloulopoulos2009, Opitz2013} and the Brown--Resnick \citep{Brown1977,Kabluchko2009} max-stable models, and \citet{Oesting2015} introduced a bivariate Brown--Resnick max-stable process to jointly model the spatial observations and forecasts of wind gusts in Northern Germany. In this paper, we propose a new class of multivariate max-stable processes that extends the Reich--Shaby model \citep{Reich2012} to the multivariate setting, and that is suitable for studying the spatial and cross-dependence structures of multiple max-stable random fields within an intuitive and computationally convenient hierarchical tree-based framework.

In contrast to the standard spatial processes based on the Gaussian distribution, the computationally demanding nature of the likelihood function for max-stable processes has hampered their use in high-dimensional settings within both frequentist and Bayesian frameworks \citep{Huser2013,Bienvenue2014,Castruccio.etal:2016,Thibaud.etal:2016}. In the Bayesian context, \citet{Thibaud.etal:2016} showed how the Brown--Resnick max-stable process may be fitted using a well-designed Markov chain Monte Carlo (MCMC) algorithm; however, this remains excessively expensive in high dimensions. From a computational perspective, it is convenient to relax the max-stable structure by assuming conditional independence of the extreme observations given an unobserved latent process \citep{Casson1999, Cooley2007, Davison2012b, Opitz.etal:2018}. This significantly facilitates Bayesian and likelihood-based inference and may be helpful for estimating marginal distributions by borrowing strength across locations. Unfortunately, when the latent process is Gaussian, the resulting dependence structure lacks flexibility and cannot capture strong extremal dependence. The Reich--Shaby model \citep{Reich2012} on the other hand possesses a conditional independence representation given some latent $\alpha$-stable random effects, and is jointly max-stable. Other popular max-stable processes do not possess such a convenient hierarchical characterization. In this paper, we generalize the Reich--Shaby process for the modeling of multivariate spatial extremes by assuming a nested, tree-based, $\alpha$-stable latent structure, which allows us to maintain a moderate computational burden.

In our proposed model, each variable is described using a Reich--Shaby spatial process, and the dependence among variables at different locations may be expressed in terms of an asymmetric max-mixture of multivariate nested logistic distributions \citep{Stephenson2003}. 
The global dependence structure may be represented by a tree framework, in which the ``leaves'' (i.e., the terminal nodes) correspond to the different spatial max-stable processes representing different variables of interest (e.g., pollutants), and the tree ``branches'' describe the relationships among these processes, which are grouped into clusters. The intra-cluster cross-dependence is assumed to be exchangeable and stronger than the inter-cluster cross-dependence. In principle, the underlying tree structure might involve an arbitrary number of ``layers'' (i.e., node levels) in order to describe more complicated forms of cross-dependence among the spatial processes, although more complex trees necessarily imply an increased number of latent variables and parameters, thus complicating the inference procedure. 

The remainder of this paper is organized as follows. The theoretical background for the geostatistical modeling of extremes is reviewed in \S\ref{geostatistics}. 
In \S\ref{nested}, we introduce a novel class of hierarchical max-stable processes built from nested $\alpha$-stable random effects for the joint modeling of multivariate extremes over space, and study its dependence properties. We also detail our inference procedure based on an MCMC algorithm, and describe some simulation experiments. 
In \S\ref{LA}, we use our nested max-stable model to investigate the spatial and cross-dependence structures of the concentration maxima of various air pollutants and temperature, observed at a number of sites across the Los Angeles area in California, US. \S\ref{summary_sp} concludes with some final remarks and perspectives for future research.

\section{Geostatistical modeling of extremes}\label{geostatistics} 
\subsection{Definition of max-stable processes}\label{sec:defmaxstable}
Owing to their asymptotic characterization, max-stable processes are widely used for modeling spatially-indexed block maxima data. In this section, we briefly summarize the general theory and modeling of max-stable processes in the spatial context, then we extend these processes to the multivariate spatial framework in \S\ref{nested}. 
Let $\{Y_1(\mathbf{s}), \ldots, Y_n(\mathbf{s})\}_{\mathbf{s} \in \mathcal S}$ be independent and identically distributed stochastic
processes  defined over the spatial region $\mathcal S\subset\mathbb R^2$. If there exist normalization functions ${a_n(\mathbf{s}) > 0}$, ${b_n(\mathbf{s})}\in \mathbb{R}$ such that the renormalized process of the pointwise maxima, i.e., 
\begin{equation}\label{eq:max.stable}
M_n(\mathbf{s}) = \frac{\max\{{Y_1(\mathbf{s}),\ldots , Y_n(\mathbf{s})}\}-b_n(\mathbf{s})}{a_n(\mathbf{s})}, \quad \quad \mathbf{s} \in \mathcal S,
\end{equation}
converges in the sense of finite-dimensional distributions to a process $Z^\star(\mathbf{s})$ with non-degenerate margins, as $n\rightarrow\infty$, then the limit  $Z^\star(\mathbf{s})$ is a max-stable process \citep[see, e.g.,][]{DeHaan1984}. The max-stability of $Z^\star(\mathbf{s})$ implies that there exist functions $\alpha_n(\mathbf{s})>0$ and $\beta_n(\mathbf{s})$ for all $n\in\mathbb N$, such that, for each collection of sites $\mathbf{s}_1,\ldots,\mathbf{s}_D\in \mathcal S$, $D\in\mathbb N$, the finite-dimensional distribution $G(z_1,\ldots,z_D)$ of the variables $Z^\star(\mathbf{s}_1),\ldots,Z^\star(\mathbf{s}_D)$ satisfies
$$G^n\{\alpha_n(\mathbf{s}_1)z_1+\beta_n(\mathbf{s}_1),\ldots,\alpha_n(\mathbf{s}_D)z_D+\beta_n(\mathbf{s}_D)\}=G(z_1,\ldots,z_D).$$ 
The limit max-stable process $Z^\star(\mathbf{s})$ may be used to represent monthly maxima of daily measurements for a specific air pollutant observed at various locations $\mathbf{s}$ within the study region $\mathcal S$. For each location $\mathbf{s} \in \mathcal S$, the Extremal Types Theorem \citep[see, e.g.,][Chapter 3]{Coles:2001} implies that the random variable $Z^\star(\mathbf{s})$ follows the generalized extreme-value (GEV) distribution with location $\mu(\mathbf{s})\in\mathbb R$, scale $\sigma(\mathbf{s})>0$, and shape $\xi(\mathbf{s})\in\mathbb R$ parameters. 
To disentangle marginal and dependence effects, it is convenient to standardize $Z^\star(\mathbf{s})$ as
\begin{equation}\label{eq:transf}
Z(\mathbf{s})=\left\{1+\xi(\mathbf{s}) \frac{Z^\star(\mathbf{s})-\mu(\mathbf{s})}{\sigma(\mathbf{s})}\right\}^{1/\xi(\mathbf{s})}.
\end{equation}
Thus, we obtain a residual, \emph{simple}, max-stable process $Z(\mathbf{\mathbf{s}})$, which is characterized by unit Fr\'{e}chet marginal distributions, i.e., $\Pr\{Z(\mathbf{s}) \leq z\} = \exp(-1/z)$, $z>0$, for all $\mathbf{s} \in \mathcal S$, corresponding to the case $\mu(\mathbf{s})=\sigma(\mathbf{s})=\xi(\mathbf{s})=1$. In practice, data are collected at a finite set of locations, and the joint distribution of $Z(\mathbf{s})$ at $\mathbf{s}_1,\ldots,\mathbf{s}_{D}\in \mathcal S$ is necessarily a multivariate extreme-value distribution that may be expressed as
\begin{equation}\label{eq:max.V}
\Pr\{Z(\mathbf{s}_1)\leq z_1, \ldots,Z(\mathbf{s}_{D})\leq z_D\}=\exp\left\{-V(z_1, \ldots,z_D)\right\},  \quad z_1,\ldots,z_D>0,
\end{equation}
where $V(z_1,\ldots,z_D)$ is the associated \emph{exponent function} containing information about the spatial dependence of the maxima. By max-stability, we can verify that $V$ is homogeneous of order $-1$, i.e., $V(tz_1,\ldots,tz_D)=t^{-1}V(z_1,\ldots,z_D)$, $t>0$; moreover, because of the unit Fr\'echet margins in \eqref{eq:max.V}, we have $V(z,\infty,\ldots,\infty)=1/z$ for any permutation of the arguments. In particular, in the case of independence between $Z(\mathbf{s}_1), \ldots,Z(\mathbf{s}_{D})$ we have $V(z_1, \ldots,z_D) = \sum_{d=1}^D z_d^{-1}$; in the case of perfect positive dependence, we have $V(z_1, \ldots,z_D) = \underset{1\leq d \leq D}{\max} z_d^{-1}$. The pairwise extremal coefficient $\theta(\mathbf{s}_i,\mathbf{s}_j)=V(1,1)\in[1,2]$ (proposed by \citealp{Smith1990}; see also \citealp{Schlather2003}), where $V$ is here restricted to  sites $\mathbf{s}_i$ and $\mathbf{s}_j$, is a measure of the extremal dependence between the variables $Z(\mathbf{s}_i)$ and $Z(\mathbf{s}_{j})$, $i,j=1,\ldots,D$. Perfect dependence corresponds to $\theta(\mathbf{s}_i,\mathbf{s}_j)=1$, and $\theta(\mathbf{s}_i,\mathbf{s}_j)=2$ leads to complete independence.

\subsection{Spectral representation of max-stable processes} \label{sec:3.3}
The construction of simple max-stable processes may be described following the spectral characterization of \citet{DeHaan1984} (see also \citealp{Penrose1992}, and \citealp{Schlather2002}). Let $W_i(\mathbf{s})$, $i\in\mathbb N$, be independent realizations of a stochastic process $W(\mathbf{s})$ such that $\textrm{E}[\sup_{\mathbf{s}\in \mathcal S}W(\mathbf{s})]<\infty$ with $\textrm{E}[W(\mathbf{s})] = 1$ for all $\mathbf{s} \in \mathcal S$, and $\zeta_i\in\Pi$ be points of the Poisson process $\Pi$ with intensity $\textrm{d}\zeta/\zeta^{2}$ on $\left(0,\infty\right)$. Then, the process constructed as
\begin{equation}\label{eq:spectralrep}
 Z(\mathbf{s}) = \max_{i\geq 1} \zeta_i W_i(\mathbf{s}),\qquad \mathbf{s} \in \mathcal S,
 \end{equation}
is simple max-stable; its joint distribution function may be written as $\Pr\left\{Z(\mathbf{s})\leq z(\mathbf{s}), \mathbf{s} \in \mathcal S\right\} = \exp\left(-\textrm{E}\left[\sup_{\mathbf{s}\in \mathcal S}\left\{{W(\mathbf{s})/z(\mathbf{s})}\right\}\right]\right)$, where $z(\mathbf{s})$ is any suitable function defined on $\mathcal S$. From \eqref{eq:max.V}, it follows that
$$V(z_1, \ldots ,z_D)=\textrm{E}\left[\sup_{\mathbf{s}\in \mathcal S}\left\{{W(\mathbf{s})\over z(\mathbf{s})}\right\}\right].$$
Choosing a different process $W(\mathbf{s})$ will lead to different classes of max-stable processes. Various parametric models have been proposed in the literature. The most well-known in the spatial framework are the \citet{Smith1990}, \citet{Schlather2002b}, Brown-Resnick \citep{Brown1977,Kabluchko2009}, extremal-$t$ \citep{Nikoloulopoulos2009, Opitz2013}, \cite{Reich2012}, and Tukey \citep{Xu2017} max-stable processes. For comprehensive reviews of the max-stable processes, see, e.g., \cite{Davison2012}, \citet{Ribatet2013} and \citet{Davison.etal:2018}.


\subsection{Hierarchical modeling based on $\alpha$-stable random effects} \label{hierarchical}


One strategy for constructing hierarchical extreme-value models is to assume that the random variables $Z^\star(\mathbf{s})$, which are observed at locations $\mathbf{s}_1,\ldots,\mathbf{s}_{D}\in\mathcal S$ and arise in the limit of renormalized maxima in \eqref{eq:max.stable}, are independent, conditional on some unobserved latent random effects; see \cite{Banerjee2004} for examples of similar hierarchical spatial models in the classical geostatistical literature. The latent Gaussian process approaches of \citet{Casson1999}, \citet{Cooley2007}, \citet{Davison2012b}, \citet{Apputhurai2013}, \citet{Dyrrdal.etal:2015}, and \citet{Opitz.etal:2018} aim to accurately capture marginal variation by embedding random effects into the marginal extreme-value parameters, but they lack flexibility to capture strong extremal dependence. In contrast, the Reich--Shaby model \citep{Reich2012}, which is based on latent $\alpha$-stable random effects, captures both marginal and joint spatial properties of extremes. The Reich--Shaby model is the only max-stable process proposed so far that possesses such a convenient hierarchical representation of spatial extremes in terms of some independent latent random effects. By integrating out the $\alpha$-stable random effects, the process resulting from the Reich--Shaby construction is max-stable and its finite-dimensional distributions are a max-mixture of multivariate logistic distributions, with the weights depending on spatial information; this allows the joint dependence structure of the spatial extremes to be described using a convenient hierarchical framework.

The \citet{Reich2012} model has been applied to various datasets and studied in different contexts. \citet{Reich2014} used this model to study extremal dependence on temperature time series; \citet{Stephenson2015} analyzed fire danger data from about 17,000 sites using compactly supported kernel functions; and \citet{Castruccio.etal:2016} compared the numerical performance of this model estimated using composite likelihood methods. 

In this section, we detail the construction and properties of the Reich--Shaby model. Let $U(\mathbf{s})\stackrel{\textrm{i.i.d.}}{\sim}\exp(-z^{-1/\alpha})$, $z>0$, $0<\alpha\leq1$, be a random noise process accounting for small-scale, non-spatial variations in the data (such as measurement errors), called the nugget effect in classical geostatistics. Furthermore, let $\vartheta\left(\mathbf{s}\right)=\left\{ \sum_{l=1}^{L}A_{l}\omega_{l}\left(\mathbf{s}\right)^{1/\alpha}\right\}^\alpha$, $\mathbf{s}\in\mathcal S\subset\mathbb{R}^2$, be a smooth spatial process, where the kernels $\omega_{l}\left(\mathbf{s}\right)\geq0$, $l=1,\ldots,L$, are deterministic spatial profiles (or weights) such that $\sum_{l=1}^{L}\omega_{l}\left(\mathbf{s}\right)=1$ for any location $\mathbf{s}\in \mathcal S$, and the $A_l$s are independent random amplitudes following the positive $\alpha$-stable distribution, i.e., $A_l\stackrel{\textrm{i.i.d.}}{\sim} \textrm{PS}(\alpha)$. 
Although the $\alpha$-stable density function, denoted $f_{\textrm{PS}}(a;\alpha)$, does not have an explicit form for $\alpha \in (0,1)$, its Laplace transform may be expressed as  
\begin{equation}\label{laplace}
\textrm{E}(e^{-tA_l})=\int_{0}^{\infty}e^{-ta}f_{\textrm{PS}}(a;\alpha) \textrm{d} a =e^{-t^{\alpha}}, \; t\geq 0.
\end{equation} 
The max-stable process proposed by  \citet{Reich2012} with unit Fr\'{e}chet margins is defined as $Z(\mathbf{s})= U(\mathbf{s})\vartheta(\mathbf{s})$ over the region $\mathcal S$. 
A spatial process with GEV margins can be obtained by inverting the transformation \eqref{eq:transf}. 
Therefore, a hierarchical model marginalized over the random noise $U(\mathbf{s})$ may be written as 
\begin{align*}
Z^\star(\mathbf{s})|A_1,\ldots, A_L&\stackrel{\textrm{ind}}{\sim}\textrm{GEV}\{\mu^\star(\mathbf{s}), \sigma^\star(\mathbf{s}), \xi^\star(\mathbf{s})\}, \label{eq:RShierarchy}\\
A_1,\ldots, A_L&\stackrel{\textrm{i.i.d.}}{\sim}\mathbf{\textrm{PS}}(\alpha), \nonumber
\end{align*}
where $\mu^\star(\mathbf{s})=\mu(\mathbf{s})+\sigma(\mathbf{s})/\xi(\mathbf{s})\{\vartheta(\mathbf{s})^{\xi(\mathbf{s})}-1\}, \sigma^\star(\mathbf{s})=\alpha\sigma(\mathbf{s})\vartheta(\mathbf{s})^{\xi(\mathbf{s})}$, and $\xi^\star(\mathbf{s})=\alpha\xi(\mathbf{s})$. The latent random effects $A_1,\ldots, A_L$ induce max-stable spatial dependence; this specific hierarchical construction ensures that the joint distribution of $Z^\star(\mathbf{s})$ is max-stable with ${\rm GEV}\{\mu(\mathbf{s}),\sigma(\mathbf{s}),\xi(\mathbf{s})\}$ margins when the random effects $A_1,\ldots,A_L$ are integrated out.

Furthermore, the residual unit Fr\'echet spatial process $Z(\mathbf{s})$, observed at $D$ locations $ \mathbf{s}_1, \ldots, \mathbf{s}_D$, is distributed according to~\eqref{eq:max.V} with the exponent function
\begin{equation}\label{eq:RS}
V(z_1, . . . ,z_D)= \sum_{l=1}^{L}\left[\sum_{d=1}^{D}\left\{ {z_{d}\over \omega_{l}(\mathbf{s}_{d})}\right\} ^{-1/\alpha}\right]^{\alpha}, \quad \quad z_1,\ldots,z_D>0,
\end{equation}
which corresponds to a max-mixture of independent random vectors $(Z_{l;1},\ldots,Z_{l;D})^{\top}$, $l=1,\ldots,L$, that are distributed according to the logistic multivariate extreme-value distribution \citep{Gumbel1960b}, i.e., $\{Z(\mathbf{s}_1),\ldots,Z(\mathbf{s}_D)\}^\top=\left\{\max_{l=1,\ldots, L}\omega_{l}(\mathbf{s}_1) Z_{l;1}, \ldots,\max_{l=1,\ldots, L}\omega_{l}(\mathbf{s}_D)Z_{l;D}\right\}^\top$. This makes the link with the general spectral representation \eqref{eq:spectralrep} \citep[see also][]{Reich2012}. Because of the asymmetry introduced by the weights $\omega_{l}(\mathbf{s})$, the model \eqref{eq:RS} is
closely related to the asymmetric logistic model introduced by \cite{Tawn1990}. Although other kernels are possible, \citet{Reich2012} proposed using the isotropic Gaussian density function  
\begin{equation}\label{eq:kernel}
g_l(\mathbf{s}) ={1\over 2\pi\tau^2}\exp\left\{-{1\over 2\tau^2}(\mathbf{s}-\mathbf{v}_l)^\top(\mathbf{s}-\mathbf{v}_l)\right\},\quad l=1,\ldots,L,
\end{equation}
 with a bandwidth (i.e., spatial range parameter) $\tau>0$ and fixed spatial knots $ \mathbf{v}_1, \ldots,\mathbf{v}_L $ distributed over the domain $\mathcal S$. We rescale the kernels to ensure that they sum to one at each location, i.e., $\omega_l(\mathbf{s}) = g_l(\mathbf{s})\left\{\sum_{l=1}^L g_l(\mathbf{s})\right\}^{-1}$, so that the margins of $Z(\mathbf{s})$ are unit Fr\'echet.

In the following section, we explain how to extend this hierarchical max-stable construction for the joint modeling of maxima from multiple variables observed over space.
\section{Nested multivariate max-stable processes} \label{nested}

\subsection{Tree-based construction of multivariate max-stable processes} \label{sec:3.1}
We generalize the univariate Reich--Shaby process, described in \S\ref{hierarchical}, in order to capture the complex spatial and cross-dependence structures among the maxima of several variables observed at multiple sites. In our proposed multivariate spatial process, the univariate spatial margins follow the Reich--Shaby model, and interact with each other by assuming a nested tree-based structure for their latent $\alpha$-stable random effects.  

\begin{figure}[t!]
\begin{center}
\includegraphics[scale=0.75]{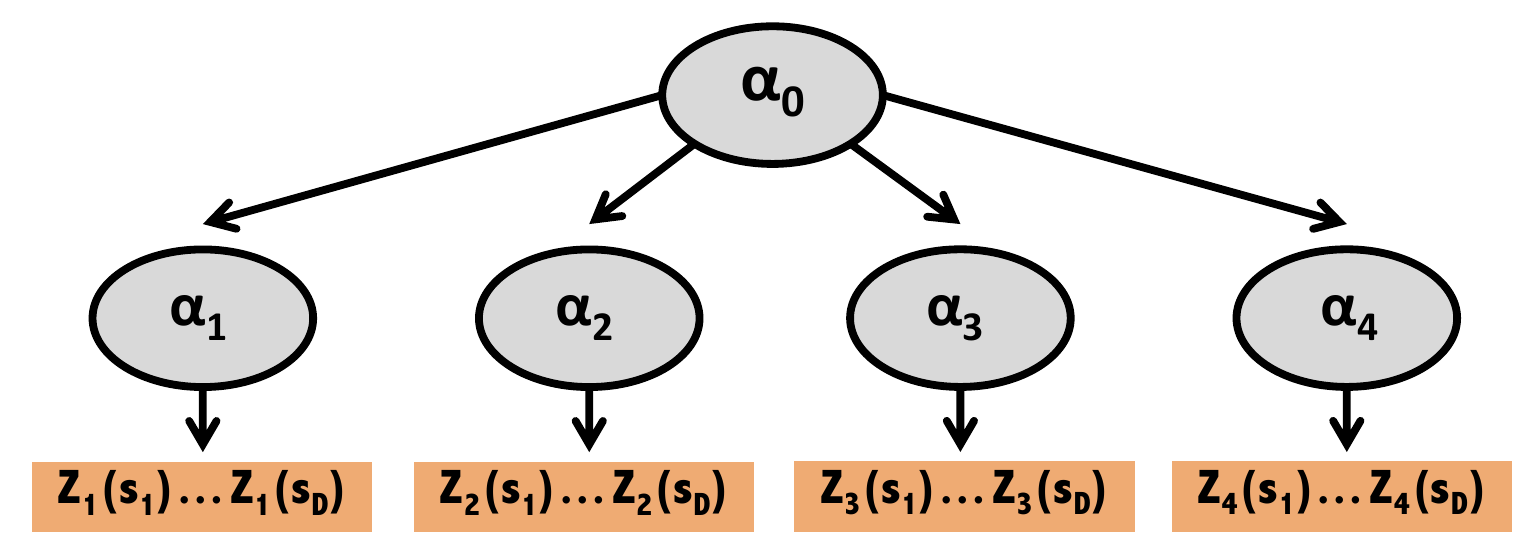}
\caption{\footnotesize{Example of simple tree structure, summarizing the extremal dependence of the process $\mathbf{Z}(\mathbf{s})=\{Z_{1}(\mathbf{s}),Z_{2}(\mathbf{s}),Z_{3}(\mathbf{s}),Z_{4}(\mathbf{s})\}^\top$. The spatial dependence of the variables $Z_{k}(\mathbf{s}), \; k=1,\ldots,K=4,$ observed at locations $\mathbf{s}_1,\ldots,\mathbf{s}_D$ is summarized by the product $\alpha_0\alpha_k$. The cross-dependence between the variables $Z_{k_1}(\mathbf{s}_i)$ and $Z_{k_2}(\mathbf{s}_j)$, $k_1\neq k_2$, is summarized by the parameter $\alpha_0$. The number of latent $\alpha$-stable random variables involved in this model is equal to the number of upper tree nodes (excluding the terminal nodes) multiplied by the number of basis functions, $L$; here, there are $5L$ latent variables.}}\label{fig:tree_2layers}
\end{center}
\end{figure}

For illustration purposes, we first define the two-layer nested multivariate max-stable process, and then gradually extend it to multilayer tree structures. Analogously to the hierarchical construction detailed in \S\ref{hierarchical}, for each $k=1,\ldots,K$, let $U_k(\mathbf{s})\stackrel{\textrm{i.i.d.}}{\sim}\exp\{-z^{-1/(\alpha_k\alpha_0)}\}$, $z>0$, denote a random Fr\'echet noise process controlled by the product of the two parameters $\alpha_k,\alpha_0\in(0,1]$; and let $\vartheta_k(\mathbf{s})=\left\{\sum_{l=1}^L A_{k;l}A_{0;l}^{1/\alpha_k}\omega_{k;l}(\mathbf{s})^{1/(\alpha_k\alpha_0)}\right\}^{\alpha_k\alpha_0}$ be a smooth spatial process, where the terms $\omega_{k;l}\left(\mathbf{s}\right)\geq0$ are $L$ kernel basis functions representing deterministic weights, such that $\sum_{l=1}^{L}\omega_{k;l}\left(\mathbf{s}\right)=1$ for any $\mathbf{s}\in \mathcal S$, and the variables $A_{k;l}$ and $A_{0;l}$, $l=1,\ldots,L$, are mutually independent (across both $k=1,\ldots,K$ and $l=1,\ldots,L$) $\alpha$-stable variables with parameters $\alpha_k$ and $\alpha_0$, respectively. Thus, $A_{k;l}\stackrel{\textrm{i.i.d.}}{\sim} \textrm{PS}(\alpha_k)\independent A_{0;l}\stackrel{\textrm{i.i.d.}}{\sim} \textrm{PS}(\alpha_0)$. 
Using the Laplace transform \eqref{laplace}, we can show that $A_{k;l}A_{0;l}^{1/\alpha_k}$ follows an $\alpha$-stable distribution with the parameter $\alpha_k\alpha_0$; therefore, each process defined as $Z_k(\mathbf{s})= U_k(\mathbf{s})\vartheta_k(\mathbf{s})$ ($k=1,\ldots,K$) is a univariate Reich--Shaby process defined over the region $\mathcal S\subset\mathbb{R}^2$, with the dependence parameter $\alpha_k\alpha_0$, kernels $\omega_{k;l}(\mathbf{s})$, $l=1,\ldots,L$, and unit Fr\'{e}chet margins. Cross-dependence among these marginal processes is induced by their shared latent variables $A_{0;1},\ldots,A_{0;L}$.
Similarly to the Reich--Shaby model, we obtain a spatial process $Z_k^\star(\mathbf{s})$ with GEV margins by inverting the transformation \eqref{eq:transf}. In this way, we write the hierarchical model marginalized over the random noise $U_k(\mathbf{s})$ as 
\begin{align*}
Z_k^\star(\mathbf{s})\mid \{A_{0;l},A_{k;l}\}_{l=1}^L&\stackrel{\textrm{ind}}{\sim} \textrm{GEV}\{\mu_k^\star(\mathbf{s}), \sigma_k^\star(\mathbf{s}), \xi_k^\star(\mathbf{s})\}, \label{eq:NRShierarchy}\\
A_{k;l}\stackrel{\textrm{i.i.d.}}{\sim}\mathbf{\textrm{PS}}(\alpha_k) &\independent A_{0;l}\stackrel{\textrm{i.i.d.}}{\sim}\mathbf{\textrm{PS}}(\alpha_0),\quad l=1,\ldots,L
\end{align*}
for  $k=1,\ldots,K$, where $\mu_k^\star(\mathbf{s})=\mu_k(\mathbf{s})+\sigma_k(\mathbf{s})/\xi_k(\mathbf{s})\{\vartheta_k(\mathbf{s})^{\xi_k(\mathbf{s})}-1\}, \sigma_k^\star(\mathbf{s})=\alpha_k\alpha_0\sigma_k(\mathbf{s})\vartheta_k(\mathbf{s})^{\xi_k(\mathbf{s})}$, and $\xi_k^\star(\mathbf{s})=\alpha_k\alpha_0\xi_k(\mathbf{s})$. 
Then, we combine the residual univariate processes $Z_{k}(\mathbf{s}), \;k=1,\ldots,K$, into the multivariate max-stable process $\mathbf{Z}(\mathbf{s})=\{Z_{1}(\mathbf{s}), \ldots,Z_{K}(\mathbf{s})\}^\top$, whose finite-dimensional distributions at $D$ locations $\mathbf{s}_{1}, \ldots, \mathbf{s}_{D}$ are expressed as~\eqref{eq:max.V} in terms of the exponent function 
\begin{equation}\label{eq:NRS}
V(\mathbf{z}_{1}, . . . ,\mathbf{z}_{K})=\sum_{l=1}^L  \left(\sum_{k=1}^K\left[\sum_{d=1}^{D}\left\{{z_{k;d}\over\omega_{k;l}(\mathbf{s}_d)}\right\}^{-1/(\alpha_k\alpha_0)}\right]^{\alpha_k}\right)^{\alpha_0},\quad z_{k;d}>0\quad \mbox{for all $k,d$.}
\end{equation}
In \eqref{eq:NRS}, $\mathbf{z}_{k}=(z_{k;1},\ldots,z_{k;D})^\top$ denotes the vector containing the maxima of the $k$th variable observed at $D$ locations, while the parameters $0<\alpha_k,\alpha_0\leq1$ ($k=1,\ldots,K$) control the spatial and cross-dependence structures. The proof of \eqref{eq:NRS} is provided in Appendix \ref{proof2}. Model \eqref{eq:NRS} corresponds to a max-mixture of $L$ independent \emph{nested} logistic max-stable distributions \citep{Tawn1990, Stephenson2003}, with the weights $\omega_{k;l}(\mathbf{s})$ introducing spatial asymmetries. This nested cross-dependence structure, represented by a simple tree in Figure \ref{fig:tree_2layers}, assumes that all the univariate processes are exchangeable. In the following, we refer to the max-stable process $\mathbf{Z}(\mathbf{s})$ with the exponent function \eqref{eq:NRS} as a two-layer nested multivariate max-stable process.

The exchangeability of such two-layer processes is not always realistic, but we overcome this limitation by generalizing the construction principle above to multilayer, partially exchangeable, tree structures based on additional nested $\alpha$-stable random effects. Because of the similarities in the ways these complex models are built, we now briefly illustrate the characterization of a three-layer nested max-stable process. We define $T$ exchangeable clusters, each comprised of $K_t$ ($t=1,\ldots,T$) max-stable processes. Thus, we have a total of $K=\sum_{t=1}^T K_t$ spatial processes. For each cluster $t=1,\ldots,T$ and variable $k=1,\ldots,K_t$, let $U_{t;k}(\mathbf{s})\stackrel{\textrm{i.i.d.}}{\sim}\exp\{-z^{-1/(\alpha_{t;k}\alpha_t\alpha_0)}\}$, $z>0$, be a random Fr\'echet noise process, and let $\vartheta_{t;k}(\mathbf{s})=\left\{\sum_{l=1}^L A_{{t;k};l}A_{t;l}^{1/\alpha_{t;k}}A_{0;l}^{1/(\alpha_{t;k}\alpha_t)}\omega_{t;k;l}(\mathbf{s})^{1/(\alpha_{t;k}\alpha_t\alpha_0)}\right\}^{\alpha_{t;k}\alpha_t\alpha_0}$ be a smooth spatial process, where $\alpha_{t;k},\alpha_t,\alpha_0\in(0,1]$ are dependence parameters, $\omega_{t;k;l}\left(\mathbf{s}\right)\geq0$ denote deterministic weights such that $\sum_{l=1}^{L}\omega_{t;k;l}\left(\mathbf{s}\right)=1$, for any $\mathbf{s}\in \mathcal S$, and $A_{t;k;l}\stackrel{\textrm{i.i.d.}}{\sim} \textrm{PS}(\alpha_{t;k})\independent A_{t;l}\stackrel{\textrm{i.i.d.}}{\sim} \textrm{PS}(\alpha_{t})\independent A_{0;l}\stackrel{\textrm{i.i.d.}}{\sim} \textrm{PS}(\alpha_0)$ are mutually independent latent $\alpha$-stable random amplitudes. Then, we set $Z_{t;k}(\mathbf{s})= U_{t;k}(\mathbf{s})\vartheta_{t;k}(\mathbf{s})$, and define the nested multivariate max-stable process with unit Fr\'echet margins as $\mathbf{Z}(\mathbf{s})=\{\mathbf{Z}_{1}^\top(\mathbf{s}), \ldots,\mathbf{Z}_{T}^\top(\mathbf{s})\}^\top$ with $\mathbf{Z}_{t}(\mathbf{s})=\{Z_{t;1}(\mathbf{s}), \ldots,Z_{t;K_t}(\mathbf{s})\}^\top$, $t=1,\ldots,T$.  Analogously to \eqref{eq:NRS}, the finite-dimensional distributions of $\mathbf{Z}(\mathbf{s})$ observed at $D$ locations $\mathbf{s}_{1}, \ldots, \mathbf{s}_{D}\in\mathcal S$ may be written as \eqref{eq:max.V} with the exponent function 
\begin{equation}\label{eq:NRS3}
V(\mathbf{z}_{1}, . . . ,\mathbf{z}_{T})=\sum_{l=1}^L  \left\{\sum_{t=1}^T\left(\sum_{k=1}^{K_t}\left[\sum_{d=1}^{D}\left\{{z_{t;k;d}\over\omega_{t;k;l}(\mathbf{s}_d)}\right\}^{-1/(\alpha_{t;k}\alpha_t\alpha_0)}\right]^{\alpha_{t;k}}\right)^{\alpha_t}\right\}^{\alpha_0}, \;\;z_{t;k;d}>0\quad \mbox{for all $t,k,d$,}
\end{equation}
where $\mathbf{z}_{t}=(\mathbf{z}_{t;1}^\top,\ldots,\mathbf{z}_{t;K_t}^\top)^\top$ and $\mathbf{z}_{t;k}=(z_{t;k;1},\ldots,z_{t;k;D})^\top$, $t=1,\ldots,T$, $k=1,\ldots,K_t$. The proof of \eqref{eq:NRS3} is provided in Appendix \ref{proof2}. As illustrated in Figure \ref{fig:tree_3layer}, the multivariate process $\mathbf{Z}(\mathbf{s})$ with \eqref{eq:NRS3} is represented graphically using a tree, whereby the terminal nodes represent the marginal max-stable processes and the upper nodes describe the cross-dependence relationships among the processes. When $T=1$ and $\alpha_{1;k} \alpha_1\equiv \alpha_k$, \eqref{eq:NRS3} reduces to the two-layer case in \eqref{eq:NRS}. Hence, \eqref{eq:NRS3} provides more flexibility than \eqref{eq:NRS} for representing complex cross-dependence structures, and we call it a three-layer nested multivariate max-stable process. 
Using the inverse mapping stemming from \eqref{eq:transf}, we obtain a three-layer nested multivariate max-stable process with arbitrary GEV margins and a hierarchical formulation:
\begin{align}
Z_{t;k}^\star(\mathbf{s})\mid \{A_{0;l},A_{t;l},A_{t;k;l}\}_{l=1}^L&\stackrel{\textrm{ind}}{\sim} \textrm{GEV}\{\mu_{t;k}^\star(\mathbf{s}), \sigma_{t;k}^\star(\mathbf{s}), \xi_{t;k}^\star(\mathbf{s})\}, \label{eq:NRS3hierarchy}\\
A_{t;k;l}\stackrel{\textrm{i.i.d.}}{\sim}\mathbf{\textrm{PS}}(\alpha_{t;k}) \independent A_{t;l}&\stackrel{\textrm{i.i.d.}}{\sim}\mathbf{\textrm{PS}}(\alpha_t) \independent A_{0;l}\stackrel{\textrm{i.i.d.}}{\sim}\mathbf{\textrm{PS}}(\alpha_0),\quad l=1,\ldots,L,\nonumber
\end{align}
where the marginal parameters satisfy $\mu_{t;k}^\star(\mathbf{s})=\mu_{t;k}(\mathbf{s})+\sigma_{t;k}(\mathbf{s})/\xi_{t;k}(\mathbf{s})\{\vartheta_{t;k}(\mathbf{s})^{\xi_{t;k}(\mathbf{s})}-1\}, \sigma_{t;k}^\star(\mathbf{s})=\alpha_0\alpha_{t}\alpha_{t;k}\sigma_{t;k}(\mathbf{s})\vartheta_{t;k}(\mathbf{s})^{\xi_{t;k}(\mathbf{s})}$, and $\xi_{t;k}^\star(\mathbf{s})=\alpha_0\alpha_{t}\alpha_{t;k}\xi_{t;k}(\mathbf{s})$. By integrating out the latent $\alpha$-stable random effects, the resulting process $Z_{t;k}^\star(\mathbf{s})$ is max-stable and has GEV margins with parameters $\mu_{t;k}(\mathbf{s})$, $\sigma_{t;k}(\mathbf{s})$, and $\xi_{t;k}(\mathbf{s})$. 


\subsection{Spatial and cross-dependence properties} \label{sec:3.2}
We now investigate the dependence properties of nested multivariate max-stable processes. In a three-layer multivariate max-stable model (see \S\ref{sec:3.1}), each marginal process $Z_{t;k}(\mathbf{s})$,  $t=1,\ldots,T$, $k=1,\ldots,K_t,$ is a Reich--Shaby process with unit Fr\'echet margins; therefore, it inherits its spatial dependence properties, which were studied in depth by \citet{Reich2012}. The product $\alpha_{t;k}\alpha_t\alpha_0$ plays the role of the dependence parameter $\alpha$ in \eqref{eq:RS} and acts as a mediator between the noise component $U_{t;k}(\mathbf{s})$ and the smooth spatial process $\vartheta_{t;k}(\mathbf{s})$. Below, we describe the cross-dependence properties of our new multivariate spatial model.

Similar to \cite{Reich2012}, we propose using the Gaussian kernel \eqref{eq:kernel} with a different bandwidth $\tau_{t;k}>0$ for each marginal process, and the spatial knots $ \mathbf{v}_1, \ldots,\mathbf{v}_L \in \mathcal S$ are fixed on a regular grid. Again, we rescale, the kernels to ensure that they sum to one at each location, i.e., $\omega_{t;k;l}(\mathbf{s}) = g_{t;k;l}(\mathbf{s})\{\sum_{l=1}^L g_{t;k;l}(\mathbf{s})\}^{-1}$, $t=1,\ldots,T$, $k=1,\ldots,K_t$, with $g_{t;k;l}$ defined similarly to $g_l$ in \eqref{eq:kernel}. In practice, we must choose a number of knots $L$ that balances computational feasibility with modeling accuracy. A too small $L$ might not be realistic and could affect subsequent inferences by artificially creating a non-stationary process \citep{Castruccio.etal:2016}, whereas a too large $L$ would significantly increase the computational burden. \citet{Reich2012} suggested fixing the number of knots, $L$, such that the grid spacing is approximately equal to or smaller than the kernel bandwidth $\tau_{t;k}$. 

To understand the cross-dependence structure of \eqref{eq:NRS3} and the meaning of its parameters, we consider the product of the dependence parameters along a specific path through the underlying tree. As mentioned above, $\alpha_{t;k}\alpha_t\alpha_0$ governs the amount of noise attributed to the univariate Reich--Shaby process in the corresponding terminal node on a given path; in contrast, $\alpha_t\alpha_0$ controls the cross-dependence among the variables belonging to the same cluster $t=1,\ldots,T$. The cross-dependence among variables belonging to distinct clusters is controlled by $\alpha_0$. Since $\alpha_t\alpha_0\leq \alpha_0$, the intra-cluster cross-dependence is always stronger than the inter-cluster cross-dependence. 

For model \eqref{eq:NRS3}, the pairwise extremal coefficient $\theta\{\mathbf{s}_i,\mathbf{s}_j;(t_1;k_1),(t_2;k_2)\}\in[1,2]$ (see \S\ref{sec:defmaxstable}) summarizes the strength of dependence between each pair of variables $\{Z_{t_1;k_1}(\mathbf{s}_i),Z_{t_2;k_2}(\mathbf{s}_j)\}^\top$, with $t_1,t_2=1,\ldots,T$, $k_1=1,\ldots,K_{t_1}$, $k_2=1,\ldots,K_{t_2}$. The variables $Z_{t_1;k_1}(\mathbf{s}_i)$ (process $k_1$ in cluster $t_1$ observed at location $\mathbf{s}_i$) and $Z_{t_2;k_2}(\mathbf{s}_j)$ (process $k_2$ in cluster $t_2$  observed at location $\mathbf{s}_j$) are perfectly dependent when $\theta\{\mathbf{s}_i,\mathbf{s}_j;(t_1;k_1),(t_2;k_2)\}=1$, and completely independent when $\theta\{\mathbf{s}_i,\mathbf{s}_j;(t_1;k_1),(t_2;k_2)\}=2$. The dependence strength increases monotonically as the value of the extremal coefficient approaches unity. Writing $\theta\{\mathbf{s}_i,\mathbf{s}_j;(t_1;k_1),(t_2;k_2)\}\equiv\theta(\mathbf{s}_i,\mathbf{s}_j)$ for simplicity, we distinguish three cases from \eqref{eq:NRS3}:
\begin{equation}\label{nl_extr}
\theta(\mathbf{s}_i,\mathbf{s}_j)= \left\{
\begin{array}{ll}
\sum_{l=1}^L \left\{\omega_{t;k;l}(\mathbf{s}_i)^{1/(\alpha_{t;k}\alpha_{t}\alpha_0)}+\omega_{t;k;l}(\mathbf{s}_j)^{1/(\alpha_{t;k}\alpha_{t}\alpha_0)}\right\}^{\alpha_{t;k}\alpha_{t}\alpha_0},&  t_1=t_2=t ,\;k_1=k_2=k,\\
\sum_{l=1}^L \left\{\omega_{t;k_1;l}(\mathbf{s}_i)^{1/(\alpha_{t}\alpha_0)}+\omega_{t;k_2;l}(\mathbf{s}_j)^{1/(\alpha_{t}\alpha_0)}\right\}^{\alpha_{t}\alpha_0},& t_1=t_2=t ,\;k_1\neq k_2,\\
\sum_{l=1}^L \left\{\omega_{t_1;k_1;l}(\mathbf{s}_i)^{1/\alpha_0}+\omega_{t_2;k_2;l}(\mathbf{s}_j)^{1/\alpha_0}\right\}^{\alpha_0}, & t_1\neq t_2,\; k_1\neq k_2.\\
\end{array}\right.
\end{equation}
Hence, as the two sites get closer to each other, i.e., as $\mathbf{s}_i\to\mathbf{s}_j$, the cross-extremal coefficient reduces to
\begin{equation}\label{nl_extr2}
\theta(\mathbf{s}_i,\mathbf{s}_j)\to \left\{
\begin{array}{ll}
2^{\alpha_{t;k}\alpha_{t}\alpha_0},&  t_1=t_2=t ,\;k_1=k_2=k,\\
2^{\alpha_{t}\alpha_0},& t_1=t_2=t ,\;k_1\neq k_2,\\
2^{\alpha_0}, & t_1\neq t_2,\; k_1\neq k_2.\\
\end{array}\right.
\end{equation}
Equation \eqref{nl_extr2} clearly confirms that intra-cluster cross-dependence is stronger than inter-cluster cross-dependence. Moreover, the nugget effect is evident when we notice that $2^{\alpha_0}>1$ for all values of $\alpha_0\in(0,1]$. Figure \ref{fig:ext} illustrates these pairwise dependence properties and shows realizations of the three-layer nested max-stable model with exponent function \eqref{eq:NRS3} and the underlying tree structure displayed in Figure \ref{fig:tree_3layer}, for specific values of the dependence parameters. The pairwise extremal coefficients shown in the top row confirm that spatial dependence for each individual process is stronger than cross-dependence, and that intra-cluster cross-dependence is stronger than inter-cluster cross-dependence. The realizations displayed in the bottom row show that strong cross-dependence between two distinct variables may result in co-localized spatial extremes.

\subsection{Inference and numerical experiments}  \label{inference_sp}
Parameter estimation for this type of model may be performed within a Bayesian framework by implementing a standard Metropolis--Hastings Markov chain Monte Carlo (MH-MCMC) algorithm \citep{Hastings1970}, which takes advantage of the hierarchical formulation \eqref{eq:NRS3hierarchy}; see, e.g., \citet{Reich2012} and \cite{Apputhurai2013} for applications to precipitation extremes. In the Supplementary Material, we detail the implementation of the MH-MCMC algorithm, which draws approximate samples from the posterior distributions of the dependence parameters $\alpha_{t;k},\alpha_t,\alpha_0$, and $\tau_{t;k}$, where $t=1,\ldots,T$, $k=1,\ldots,K_t$, for the nested max-stable model \eqref{eq:NRS3}. Essentially, the dependence parameters $\alpha_{t;k}, \alpha_t, \alpha_0$, and $\tau_{t;k}$ are updated successively (one by one) at each iteration $r=1,\ldots,R$ of the algorithm by generating candidate values $\alpha^{(\textrm{c})}_{t;k}, \alpha^{(\textrm{c})}_t, \alpha^{(\textrm{c})}_0$, and $\tau^{(\textrm{c})}_{t;k}$, computing the corresponding candidate process $\vartheta^{(\textrm{c})}_{t;k}$, and finally accepting such candidate values with a certain probability that ensures theoretical convergence of the Markov chain to its stationary posterior distribution. For simplicity, we assume here that the prior distributions of the parameters $\alpha_{t;k}, \alpha_{t}$, and $\alpha_{0}$ are non-informative ${\rm Unif}(0,1)$, and that the range parameters $\tau_{t;k}$ have prior distribution equal to $0.5 h_{\max}\times{\rm Beta}(2,5)$ as suggested by \citet{Quentin2016}, where $h_{\max}$ denotes the maximum distance between stations (representing the ``diameter'' of the region of study). This slightly informative prior distribution guides and stabilizes the estimation of the range parameters, whose posterior distribution can sometimes be very right-skewed, but it should not have an important impact on the estimated model for a large $D$, i.e., when there are many monitoring stations. 

\begin{figure}[t!]
\begin{center}
\includegraphics[width=0.6\linewidth]{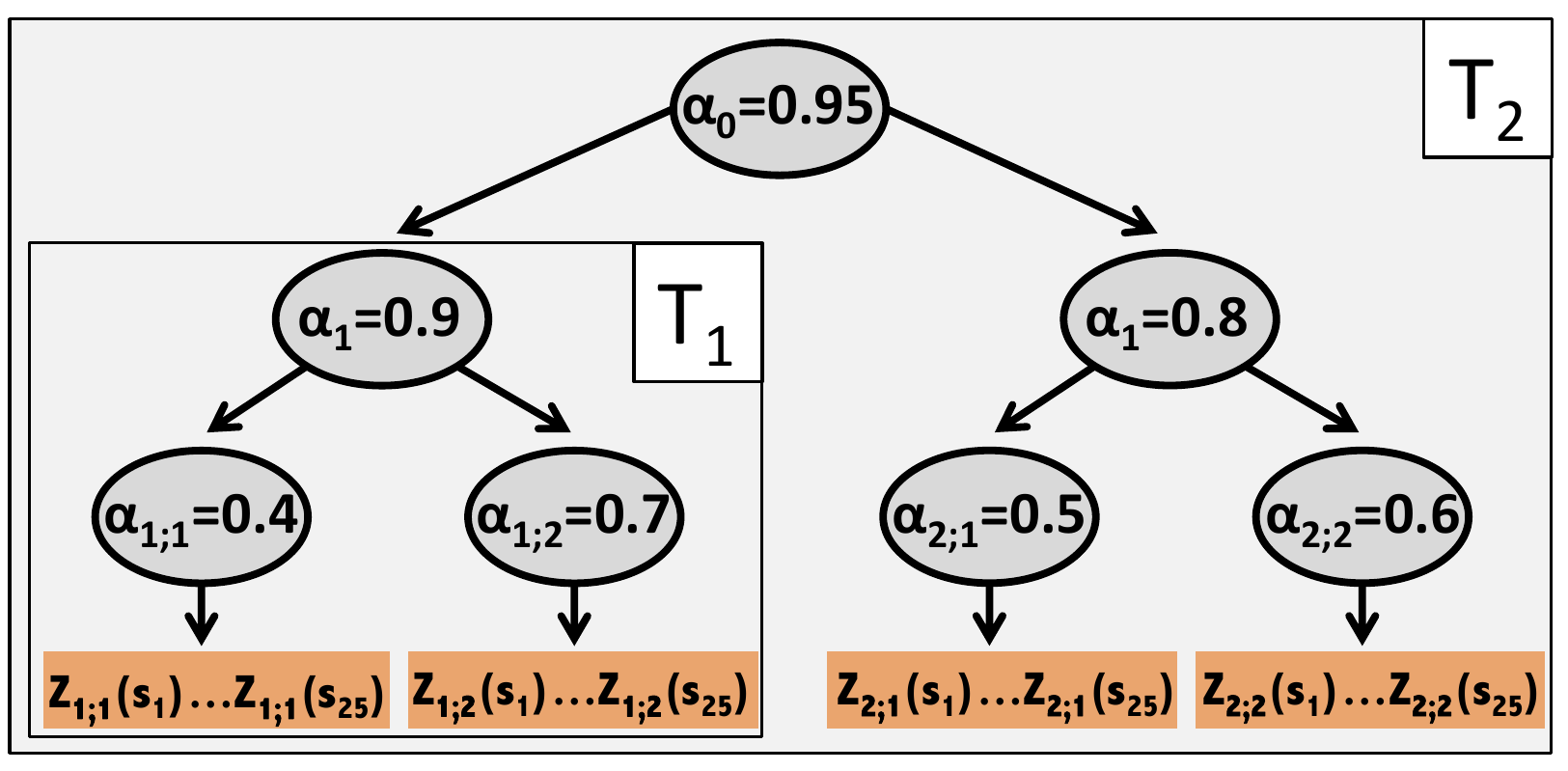}
\caption{\footnotesize{Dependence structure configurations used in our simulation study based on the nested multivariate max-stable process with  two layers (tree $T_1$) for $T=2$ and three layers (tree $T_2$) for $T=4$.}}\label{fig:configuration2}
\end{center}
\end{figure}

\begin{figure}[t!]
\begin{center}
\includegraphics[width=\linewidth]{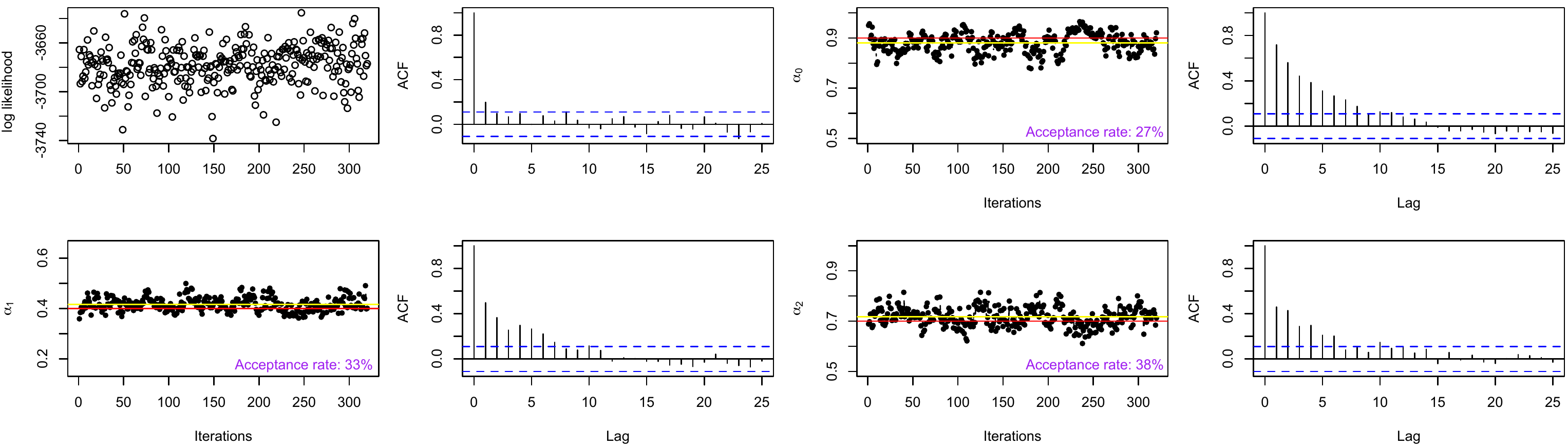}
\caption{\footnotesize{Trace-plots ($1^{\mbox{\tiny st}}$ and $3^{\mbox{\tiny rd}}$ columns) and the corresponding autocorrelation functions ($2^{\mbox{\tiny nd}}$ and $4^{\mbox{\tiny th}}$ columns) for the subchains of the log-likelihood values (top left) and the dependence parameters $\alpha_0$ (top right), $\alpha_1$ (bottom left), and $\alpha_2$ (bottom right), obtained by using the MH-MCMC algorithm to estimate model~\ref{eq:NRS} (tree structure $T_1$ in Figure \ref{fig:configuration2}), using $R=10^6$ iterations, and thinned by a factor 2500 after removing a burn-in of $R/5$ iterations. True and posterior medians are indicated by red and yellow lines, respectively. Effective sample sizes are equal to $104$ ($\alpha_0$), $595$ ($\alpha_1$), and $907$ ($\alpha_2$).}}\label{fig:output_2layers}
\end{center}
\end{figure}

\begin{figure}[t!]
\begin{center}
\includegraphics[width=\linewidth]{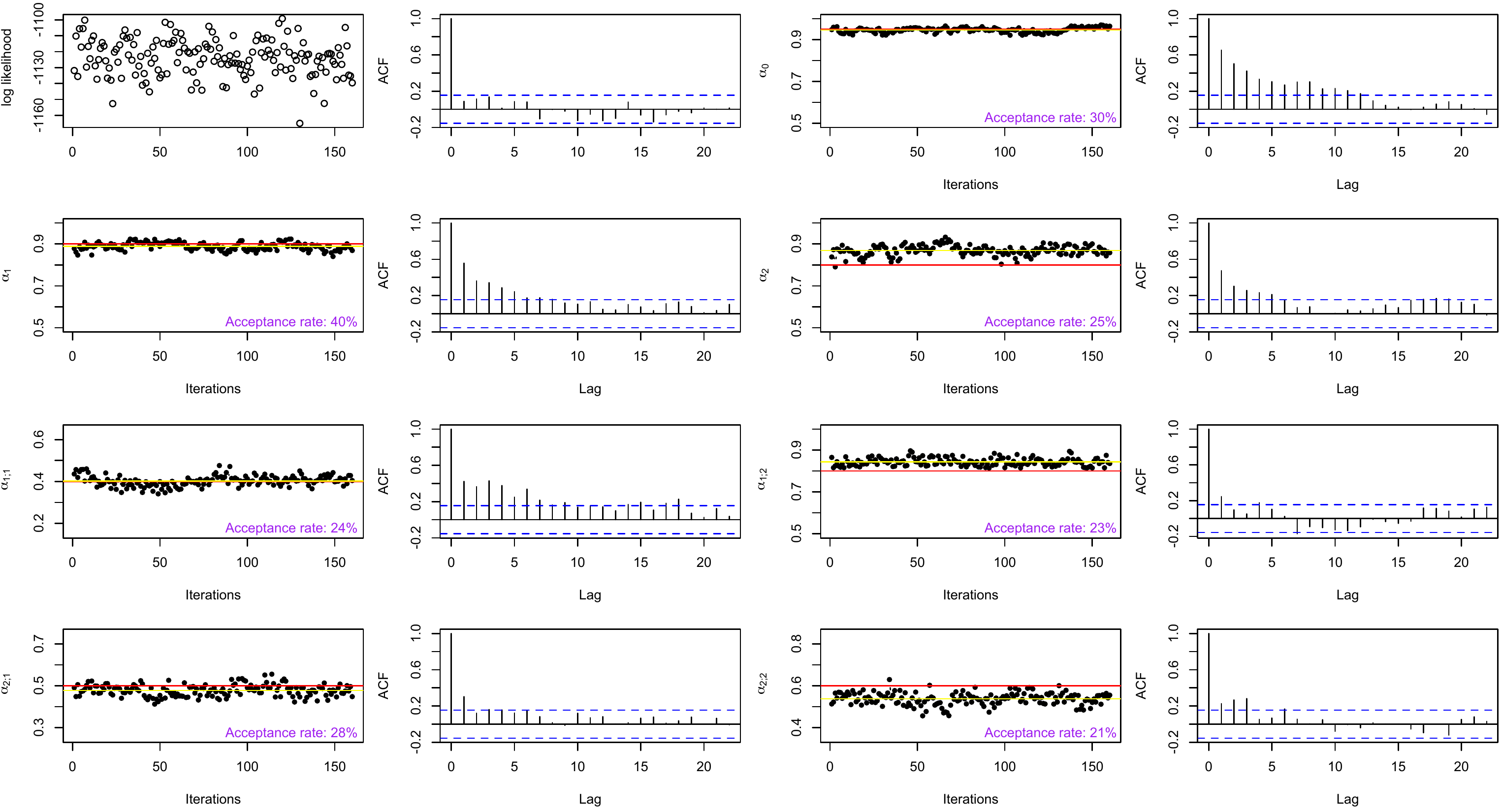}
\caption{\footnotesize{Trace-plots ($1^{\mbox{\tiny st}}$ and $3^{\mbox{\tiny rd}}$ columns) and the corresponding autocorrelation functions ($2^{\mbox{\tiny nd}}$ and $4^{\mbox{\tiny th}}$ columns) for the subchains of the log-likelihood values (top left) and the dependence parameters $\alpha_0$ (top right), $\alpha_1$ ($2^{\mbox{\tiny nd}}$ row left), $\alpha_2$ ($2^{\mbox{\tiny nd}}$ row right), $\alpha_{1;1}$ ($3^{\mbox{\tiny rd}}$ row left), $\alpha_{1;2}$ ($3^{\mbox{\tiny rd}}$ row left), $\alpha_{2;1}$ (bottom left), and $\alpha_{2;2}$ (bottom right), obtained by using the MH-MCMC algorithm to estimate model~\ref{eq:NRS3} (tree structure $T_2$ in Figure \ref{fig:configuration2}), using $R=5\times10^5$ iterations, and thinned by a factor 2500 after removing a burn-in of $R/5$ iterations. True and posterior medians are indicated by red and yellow lines, respectively. Effective sample sizes are equal to $50$ ($\alpha_0$), $108$ ($\alpha_1$) and $122$ ($\alpha_2$), $262$ ($\alpha_{1;1}$), $116$ ($\alpha_{1;2}$), $544$ ($\alpha_{2;1}$), and $749$ ($\alpha_{2;2}$).}}\label{fig:output_3layers}
\end{center}
\end{figure}

In order to verify the performance of the MH-MCMC algorithm implemented for the nested multivariate max-stable process proposed in \S\ref{nested}, we conduct a simulation study and explore some diagnostics of the output. The chosen simulation design conforms to the experiments reported by \citet{Reich2012} for each univariate process $Z_{t;k}(\mathbf{s})$, $t=1,\ldots,T$, $k=1,\ldots,K_t$, so here we focus on testing the accuracy of the estimates for the cross-dependence parameters $\alpha_{t;k}$, $\alpha_t$, and $\alpha_0$. We generate data from the nested max-stable models with two nesting layers \eqref{eq:NRS} and three nesting layers \eqref{eq:NRS3}, using the tree structures $T_1$ and $T_2$, respectively, represented in Figure \ref{fig:configuration2}, with $N = 20$ independent time replicates. 
We consider different values for the dependence parameters $\alpha_{t;k}, \alpha_t$, and $\alpha_0$, fix the bandwidth $\tau_{t;k}=3$ in each individual process, and assign the marginal parameters $\mu_{t;k}(\mathbf{s})=\sigma_{t;k}(\mathbf{s})=\xi_{t;k}(\mathbf{s})=1$, corresponding to the unit Fr\'echet case. The data are simulated on a $5 \times 5$ regular grid (i.e., $D=25$) covering the domain $\mathcal S=[0,6] \times [0,6]$, with Gaussian kernels centered on the same grid points. 
The MH-MCMC output is (partly) represented in Figure~\ref{fig:output_2layers} for the two-layer case and in Figure~\ref{fig:output_3layers} for the three-layer case. We performed $R=5\times 10^5$ iterations for the two-layer case and $R=10^6$ iterations for the three-layer case, then thinned the resulting chains by a factor $2500$ after removing a burn-in of $R/5$ iterations. By integrating {\tt C++} code in {\tt R} using the package {\tt Rcpp}, the computational time was less than two days for each experiment.

\begin{figure}[t!]
\begin{center}
\includegraphics[width=\linewidth]{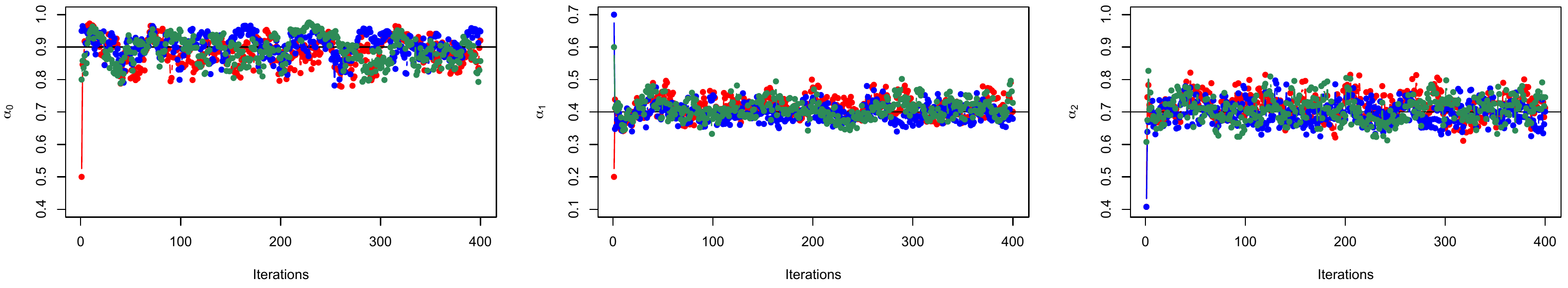}
\caption{\footnotesize{Trace-plots for the subchains corresponding to the tree structure $T_1$ represented in Figure \ref{fig:configuration2}, obtained by using the MH-MCMC algorithm with $R=10^6$ iterations choosing different starting values. The subchains shown are thinned by a factor $2500$.}}\label{fig:init}
\end{center}
\end{figure}

The trace plots of both model outputs suggest that the resulting chains are fairly stationary for all parameters. Moreover, the autocorrelation functions indicate that the subchains show relatively good mixing properties after thinning, although the effective sample sizes suggest that the parameters controlling the upper part of the hierarchy are more difficult to estimate. For example, the parameter $\alpha_0$, which mostly impacts the inter-cluster dependence, indeed appears to be the most strongly auto-correlated. Figure \ref{fig:init} shows that the MH-MCMC algorithm produces similar results from different starting values, which strongly confirms that the Markov chain indeed converges to its stationary distribution. Overall, the posterior medians of the dependence parameters coincide with the true values, indicating that our model captures both the spatial and cross-dependence structures governing the joint behavior of these max-stable processes quite well. 

\section{Multivariate spatial analysis of Los Angeles air pollution extremes} \label{LA}
\subsection{Motivation and data description}
\begin{figure}[t!]
\begin{center}
\includegraphics[width=0.6\linewidth]{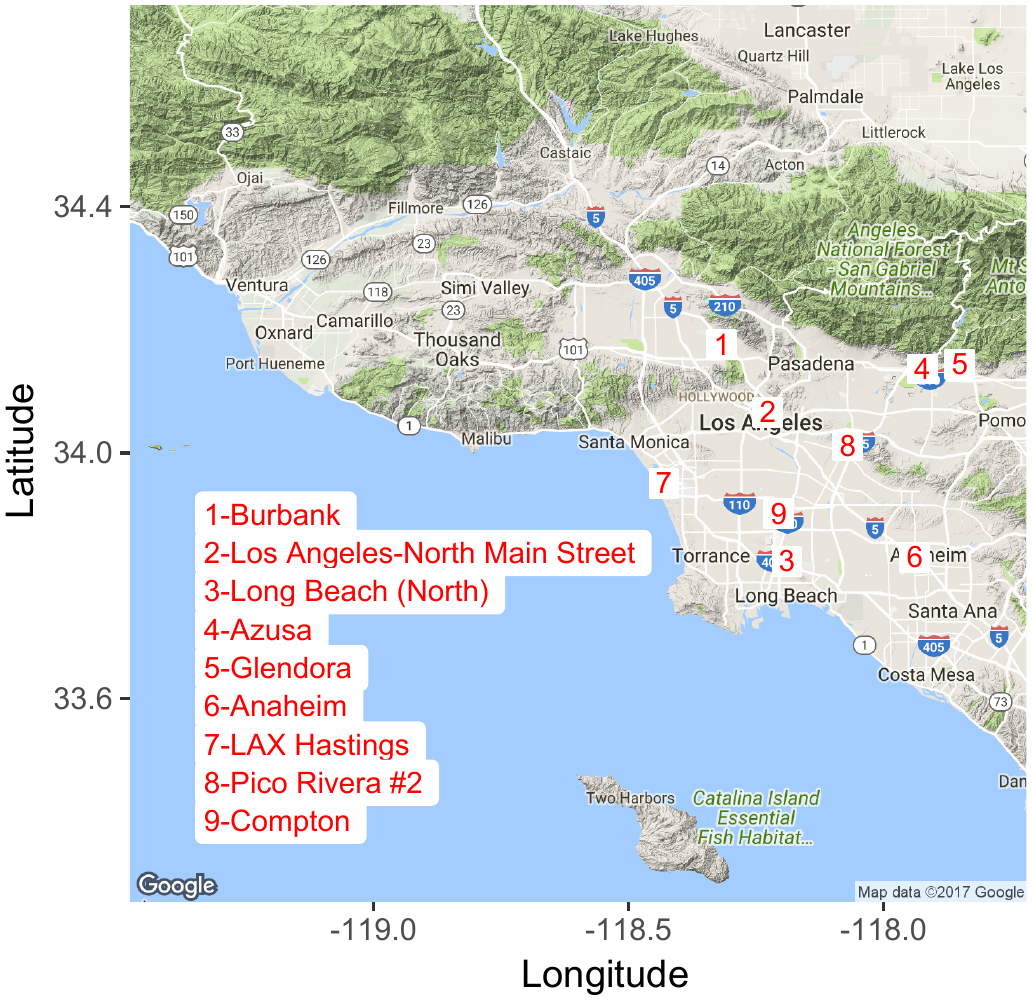}
\caption{\footnotesize{Map of Los Angeles and surrounding area with the nine sites under study indicated by numbers.
}}\label{fig:map_LA}
\end{center}
\end{figure}

High concentrations of pollution in the air can harm the human body. Current methods for assessing air pollution dangers typically consider each pollutant separately, ignoring the heightened threat of exposure to multiple air pollutants. In order to inform the public and government administrations, the US Environmental Protection Agency (EPA) and other international organizations are moving towards a multi-pollutant approach for quantifying health risks of air pollution. In this work, we investigate the extremal dependence among air pollutants and temperature jointly across space. The data consist of daily observations of carbon monoxide (CO), nitric oxide (NO), nitrogen dioxide (NO$_2$), ozone (O$_3$), and temperature (T), collected from January 2006 to December 2015, at a number of sites in the area of Los Angeles, CA; see Figure~\ref{fig:map_LA}. 
A similar study was conducted by \citet{Vettori2017}, who analyzed the extremal dependence between air pollutants and meteorological parameters at several sites in California \emph{separately}, thereby ignoring their spatial dependence. Here, we use our new methodology based on nested multivariate max-stable processes to characterize the spatial and cross-dependence structures among these variables of interest.
 
\subsection{Model fitting and diagnostics}
\begin{figure}[t!]
\begin{center}
{\centering\includegraphics[width=\linewidth]{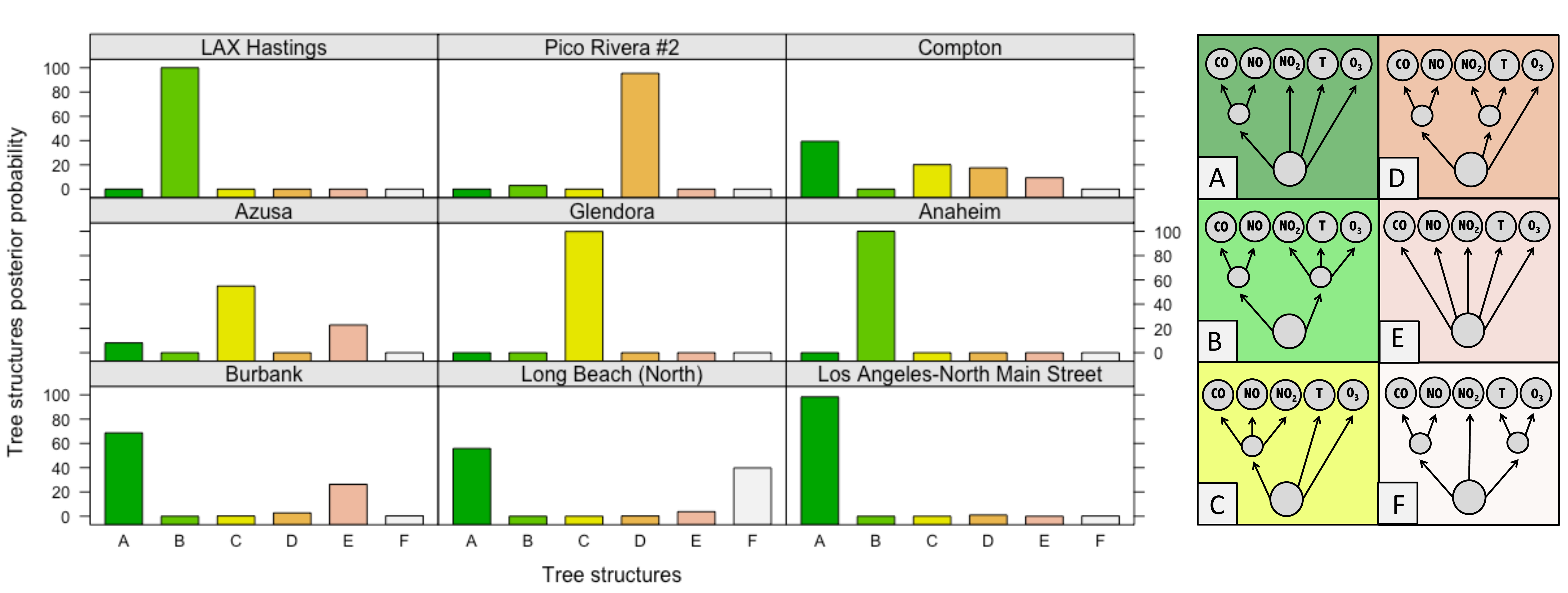}}
\caption{\footnotesize{The most frequent tree structures (right), indicated by letters A--F, identified by the TM-MCMC algorithm after $R=5000$ iterations, burn-in $R/4$ and thinning factor 5. The histograms (left) report the posterior probability associated with each tree for each site in Figure \ref{fig:map_LA} calculated according to the number of times each tree appears in the algorithm chain.
}}\label{fig:TREE_LA}
\end{center}
\end{figure}

We start by modeling the non-stationary marginal effects from the monthly maxima of each pollutant and meteorological parameter through linear regression, including trend and seasonality terms, and then we standardize these maxima data to a common unit Fr\'echet scale in order to analyze their complex multivariate dependence structure. By using the tree mixture MCMC (TM-MCMC) algorithm introduced by \citet{Vettori2017} for purely multivariate (i.e., non-spatial) data, we investigate plausible multivariate cross-dependence structures for each site separately. By exploiting reversible jump MCMC and the simple tree-based representation of the nested logistic distribution, this algorithm samples from the posterior distribution of the parameters and the tree itself. We apply the TM-MCMC algorithm with $R = 5000$ iterations and a thinning factor of $5$ to the stationary, standardized, monthly maxima time series of the variables CO, NO, NO$_2$, O$_3$, and T separately for each site displayed in Figure~\ref{fig:map_LA}. The most likely dependence structures identified by the TM-MCMC algorithm and the associated posterior probabilities are presented in Figure \ref{fig:TREE_LA}. 
The tree structures that appear most often across the chains are trees A, B, and C. We find that extreme concentrations of CO and NO are consistently grouped together in the same cluster by trees A, B and C. Moreover, the extreme concentrations of NO$_2$ are grouped with extreme concentrations of O$_3$ and high temperatures in tree B, and with extreme concentrations of CO and NO in tree C. 

\begin{figure}[t!]
\begin{center}
{\centering\includegraphics[width=\linewidth]{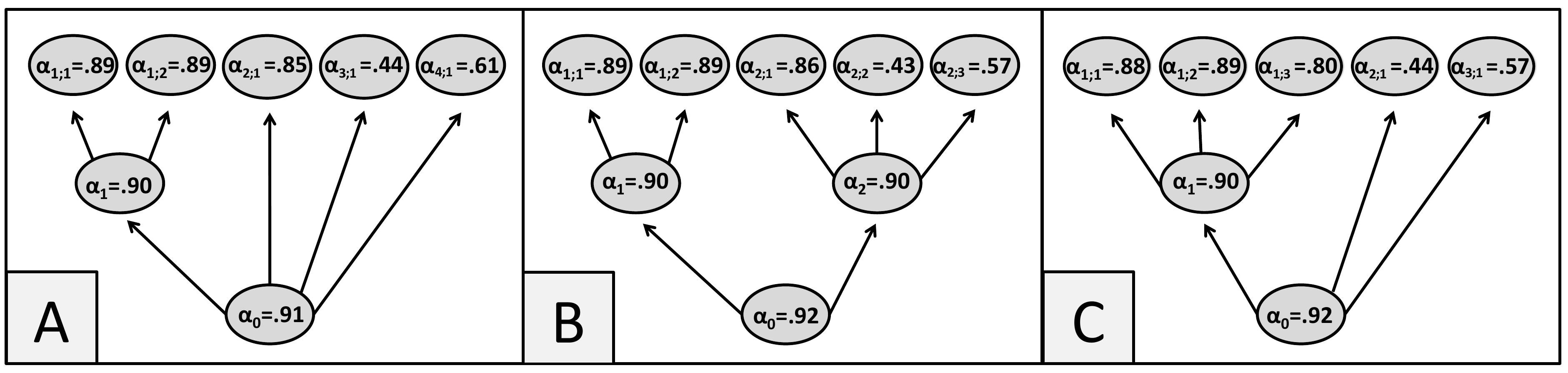}}
\caption{\footnotesize{Posterior medians of the dependence parameters $\alpha_{t;k}, \alpha_t, \alpha_0$  obtained by fitting our new three-layer nested multivariate max-stable model~\eqref{eq:NRS3} to the concentration maxima of CO, NO, NO$_2$, O$_3$, and temperature using the MCMC algorithm and assuming the tree structures A, B, or C from Figure \ref{fig:TREE_LA}. 
}}\label{fig:Posteriors}
\end{center}
\end{figure}

\begin{figure}[t!]
\begin{center}
{\centering\includegraphics[width=\linewidth]{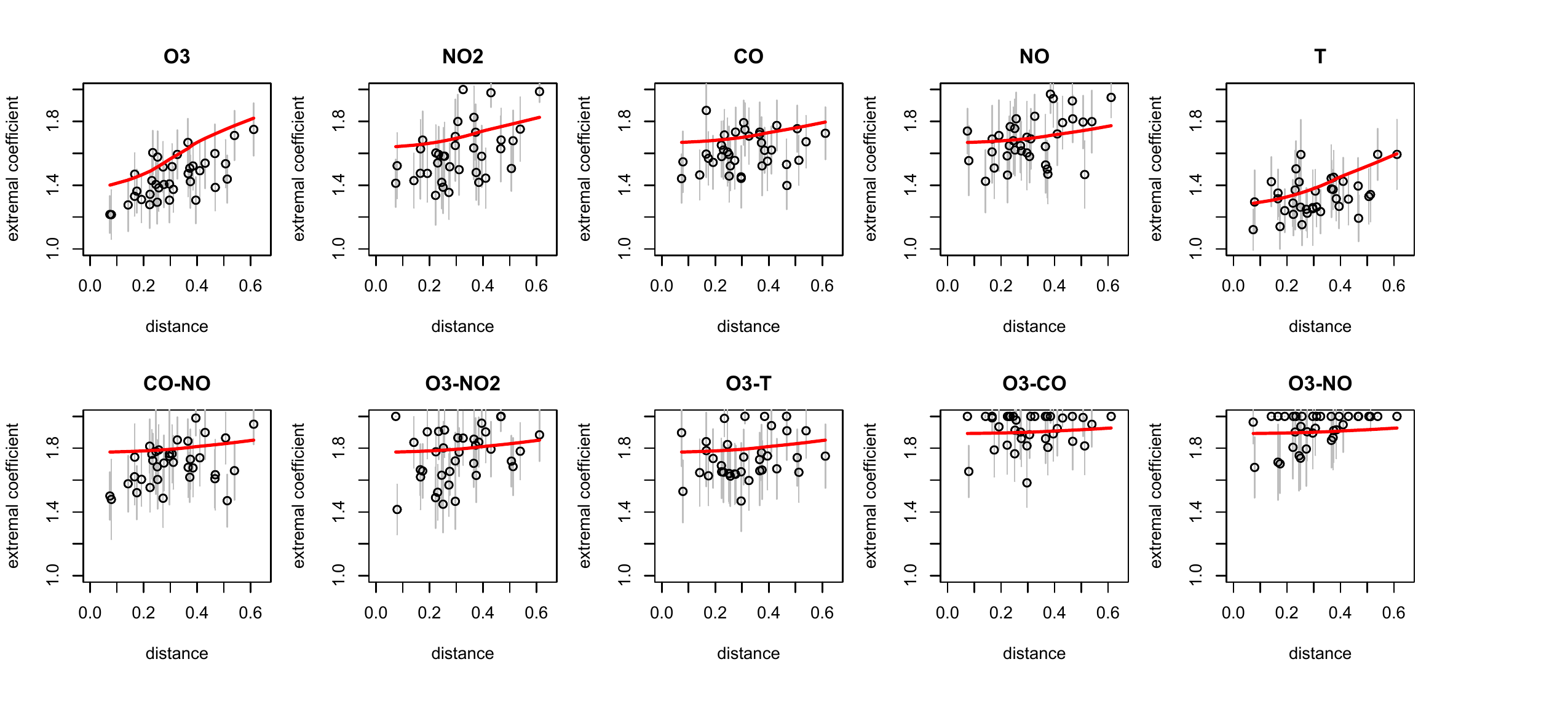}}
\caption{\footnotesize{Empirical (black dots) and model-based (red curves) pairwise extremal coefficient estimates for each individual variable (top) and some selected pairs of variables (bottom). The model-based pairwise extremal coefficients are obtained from fitting the three-layer nested multivariate max-stable model using the MCMC algorithm assuming the tree structures B. Grey vertical segments display $95\%$ confidence intervals for the empirical extremal coefficients.}}\label{fig:ext_fits}
\end{center}
\end{figure}

In Figure~\ref{fig:Posteriors}, we fit the three-layer nested multivariate max-stable model with the tree dependence structures A, B, and C; the posterior medians are reported in each node. 
The estimated parameters $\alpha_0$ take values close to $0.9$ in all three trees, indicating that the extreme concentrations of CO and NO are weakly related to the extreme concentrations of NO$_2$ and O$_3$ and to high temperatures across the Los Angeles area. Spatial dependence is strongest between the pollutants CO and NO in tree A and between CO and NO$_2$ in the case of tree C, whereas dependence among NO$_2$, O$_3$, and high temperature is strongest in the case of tree B. 
Figure \ref{fig:ext_fits} compares the empirical estimates of the pairwise extremal coefficients (see \citealp{Smith1990} and \S\ref{sec:defmaxstable}) with the pairwise extremal coefficients computed from the fitted nested multivariate max-stable model \eqref{nl_extr}, using the dependence structure of tree B. Generally, model-based estimates are fairly similar to their empirical counterparts (given that the empirical estimates are highly variable), suggesting that the fitted model is reasonable and adequately captures the complex spatial cross-dependence structure of extremes in our dataset. Considering more complex tree structures might further improve the fit. As expected, the pairwise extremal coefficient estimates computed for individual variables seem to increase with the distance between sites. Moreover, both the empirical and model-based pairwise cross-extremal coefficient estimates indicate a moderate dependence strength between variables belonging to the same cluster, such as CO and NO,  O$_3$ and NO$_2$, or O$_3$ and T, regardless of the distance between sites, whereas the variables belonging to different clusters, such as O$_3$ and CO or O$_3$ and NO, appear to be almost independent at any distances.

\subsection{Return level projections and air pollution risk assessment}
\begin{figure}[t!]
\begin{center}
{\centering\includegraphics[scale=0.46]{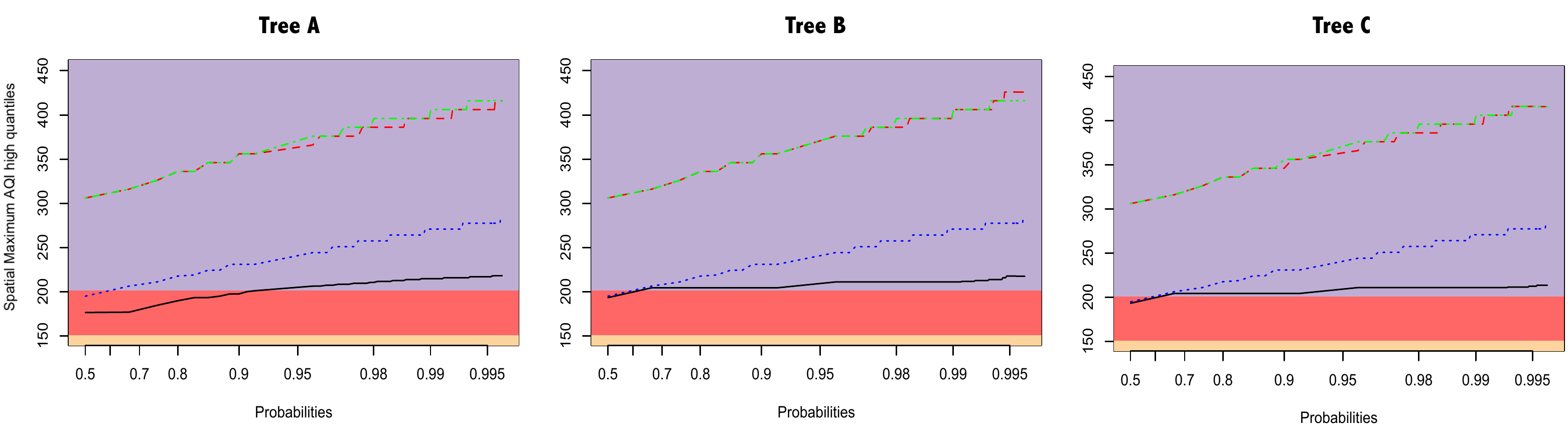}}
\caption{\footnotesize{ 
High $p$-quantiles $z_p$ computed for the spatial maximum of the largest Air Quality Index (AQI) for CO, O$_3$, and NO$_2$, setting April 2009 as baseline, obtained by fitting the nested multivariate max-stable model (black lines); by fitting the Reich--Shaby model to each individual process separately, neglecting cross-dependence (blue lines); by running the TM-MCMC algorithm to estimate the cross-dependence between variables at each site separately, neglecting spatial dependence (red lines); and by fitting the GEV distribution to each site and pollutants independently, neglecting spatial and cross-dependence structures (green lines). Results are shown underlying trees A (left), B (middle), and C (right). Probabilities are displayed on a Gumbel scale, i.e., $z_p$ is plotted against $-\log\{-\log(p)\}$. AQI categories: 0-50 satisfactory (green);  51-100 acceptable (yellow); 101-150 unhealthy for sensitive groups (orange); 151-200 unhealthy (red); $>$200 very unhealthy (purple). 
}}\label{fig:AQI_LA}
\end{center}
\end{figure}

The US EPA typically uses the Air Quality Index (AQI) to communicate air pollution risks to the public. In order to illustrate the impacts of neglecting spatial and cross-dependence structures on the AQI return level estimates, Figure~\ref{fig:AQI_LA} shows high $p$-quantiles with probabilities ranging from $p=0.5$ to $p=0.996$, considering April 2009 as the baseline, computed for the maximum AQI over the pollutants CO, O$_3$, and NO$_2$ and across space, using different models. 
Under stationary conditions, the return levels for $1$ and $20$ years roughly correspond to $p=1-1/12\approx0.917$ and $p=1-1/(12\times 20)\approx 0.996$, respectively. The AQI categories, representing different levels of health concern, are represented by different colors. 
In particular, we compare the fits of the full nested multivariate max-stable model, the Reich--Shaby model fitted to each individual process separately (i.e., ignoring cross-dependence), the nested logistic distribution fitted at each site separately using the TM-MCMC algorithm (i.e., ignoring spatial dependence), and the GEV distribution fitted to each site and pollutant independently (i.e., ignoring both spatial and cross-dependence). The AQI quantiles obtained from the posterior predictive distribution of the Reich--Shaby spatial fits, are much smaller than the ones obtained from the multivariate distribution fitted to each site separately or the GEV distribution fitted to each site and pollutants separately. Furthermore, the high quantile projections calculated based on the nested multivariate max-stable model are generally smaller than the high quantiles based on the Reich--Shaby model. Therefore, when neglecting spatial dependence or the multivariate cross-dependence among processes, the return levels calculated for the maximum of several extreme observations may be strongly overestimated. Similar results were found by \cite{Huser2016b} when they explored the effect of model misspecification in a non-stationary context. Using our proposed multivariate max-stable process based on tree A, the high quantiles $z_p$ (for the maximum AQI across all sites and pollutants) lie within the very unhealthy category for probabilities $p\geq 0.92$, indicating that at least one of the criteria pollutants under study exceeds this critical threshold at one or more of the monitoring sites approximately once every year. We obtained similar results from trees B and C.

\begin{figure}[h!]
\begin{center}
{\centering\includegraphics[width=0.89\linewidth]{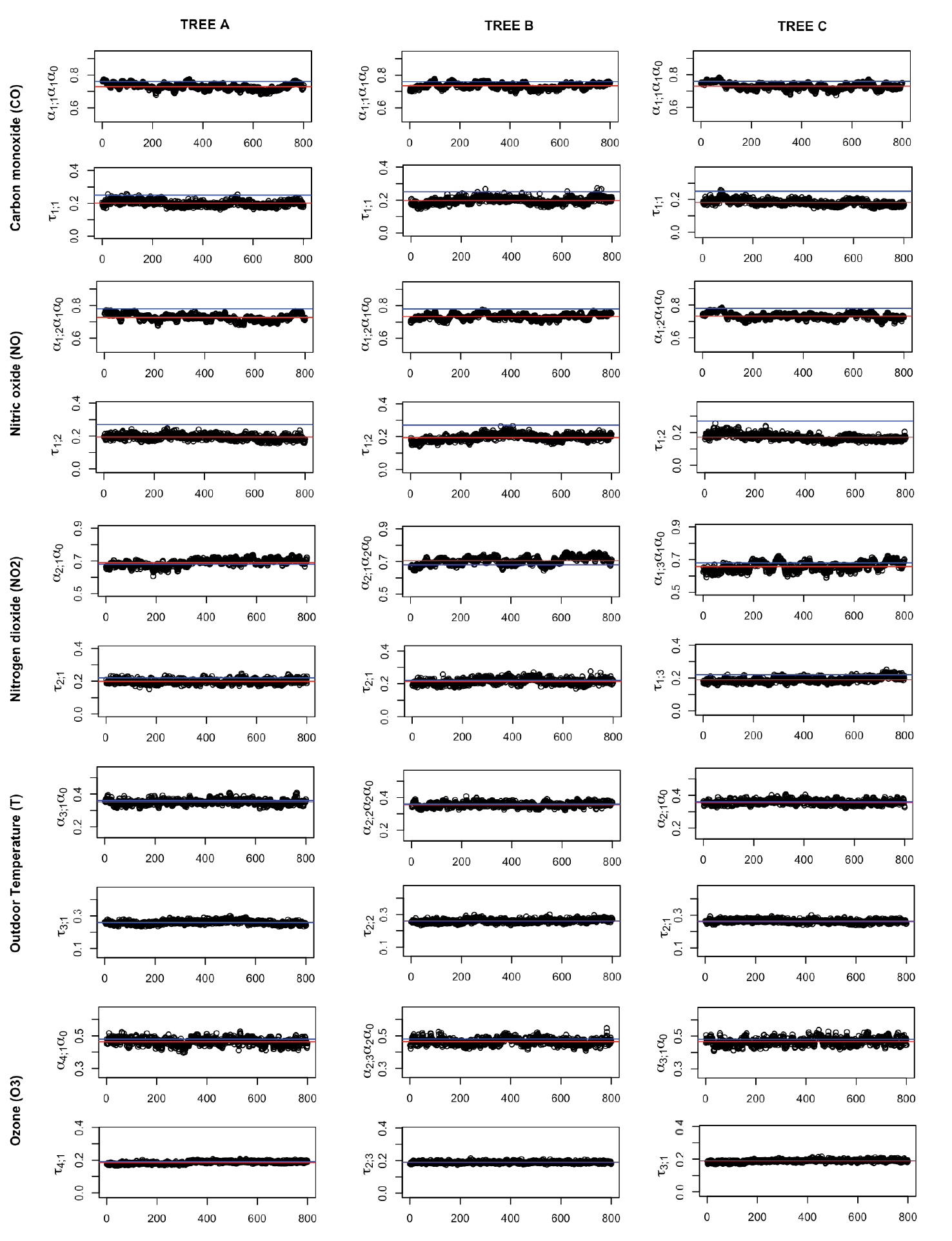}}
\caption{\footnotesize{Trace plots and posterior medians (red) for the subchains of the products of the dependence parameters $\alpha_{t;k}\alpha_t\alpha_0$ and the kernel bandwidths $\tau_{t;k}$ (for each variable $k$ in cluster $t$, see rows) obtained by fitting the nested multivariate max-stable model using the MCMC algorithm, assuming the tree structures A (left), B (middle), and C (right) in Figure \ref{fig:TREE_LA}. The corresponding posterior medians for $\alpha$ and $\tau$ obtained by fitting the Reich--Shaby model to each process separately are indicated by blue lines.}}\label{fig:RS_LA}
\end{center}
\end{figure}

To verify that the nested multivariate max-stable model provides a good marginal fit for each of the processes under study, Figure \ref{fig:RS_LA} compares the MH-MCMC algorithm output obtained from the joint fit to the posterior medians of the dependence parameters $\alpha$ and $\tau$ obtained from the Reich--Shaby model fitted to each process independently. Overall, the joint and individual models provide similar values for the marginal parameters, confirming that our approach yields sensible marginal fits, while simultaneously providing information about the cross-dependence structure. However, the processes characterized by a large dependence parameter $\alpha$ are quite noisy by nature, and therefore harder to estimate. For example, there is a slightly larger mismatch between the dependence parameter estimates obtained from the joint and individual fits for the spatial variables CO and NO. 

\section{Conclusion} \label{summary_sp}
We introduced a novel class of hierarchical multivariate max-stable processes that have the Reich--Shaby model as univariate margins, and that can capture the spatial and cross-dependence structures among extremes of multiple variables, based on latent nested $\alpha$-stable random effects. These hierarchical models may be conveniently represented by a tree structure, and the complexity of the dependence relations among the various spatial variables might be increased by adding an arbitrary number of nesting layers. 
Parameter estimation can be carried out within a Bayesian framework using a standard Metropolis--Hasting MCMC algorithm.  As shown in our simulation experiments, the dependence parameters governing the spatial dependence of individual variables and the cross-dependence among different variables can be satisfactorily identified using our proposed algorithm. 

We fitted the nested multivariate max-stable process to air pollution extremes collected in the Los Angeles area. In addition to providing good spatial marginal fits for each of the air pollutants under study, our model detects their extremal spatial cross-dependence, and takes into account the temperature extremes.  
Extreme concentrations of toxic air pollutants, such as CO, NO, NO$_2$, and O$_3$ and extremely high temperatures are weakly related across the area of Los Angeles. Furthermore, a strong cross-dependence is detected between the maxima of CO and NO, which are both pollutants released by fossil fuel combustion. Also, high concentrations of O$_3$ and high temperatures often occur simultaneously, which leads to a heightened health threat according to \citet{Kahle2015}. Modeling air pollution extremes using the proposed nested multivariate max-stable model allows us to provide sensible multi-pollutant return level estimates based on the Air Quality Index (AQI); thus, our new methodology is useful for assessing the risks associated with simultaneous exposure to several air pollutants over space, and might be used to develop future air pollution monitoring regulations. 

In order to fit the nested multivariate max-stable process, we must assume a single fixed tree structure across space. It would be interesting to investigate how we can generalize this model to account for spatially-varying tree structures. One possible starting point could be to define homogeneous subregions governed by different cross-dependence structures. 
  

\baselineskip=16pt
\bibliography{Bibliography}
\bibliographystyle{CUP}

\begin{appendix} 
\baselineskip=16pt

\section*{Appendix}

\section{Proof of \eqref{eq:NRS3}}\label{proof2}
Below, we use the Laplace transform of an $\alpha$-stable random variable \eqref{laplace}, and write the vectors of latent $\alpha$-stable random effects as $\mathbf{A}_0 = (A_{0;1}, \dots, A_{0;L})^\top,\; \mathbf{A}_t = (A_{t;1}, \dots, A_{t;L})^\top, \; \mathbf{A}_{t;k} = (A_{t;k;1}, \dots, A_{t;k;L})^\top$. The multivariate distribution function of $\mathbf{A}_0$, $\mathbf{A}_t$ or $\mathbf{A}_{t;k}$ is given by the product of the $L$  $\alpha$-stable marginal distribution functions, as $A_{0;l}$, $A_{k;l}$, and $A_{t;k;l}$ are independent for $l=1,\ldots,L$. The joint distribution function $G$ of the random vector
$\{Z_{1;1}(\mathbf{s}), \ldots, Z_{1;K_1}(\mathbf{s}), \ldots, Z_{T;1}(\mathbf{s}), \ldots, Z_{T;K_T}(\mathbf{s})\}^\top$ at locations $\mathbf{s}_{1}, \ldots, \mathbf{s}_{D}$ is 
\begin{align*}
&\Pr\{Z_{t;k}(\mathbf{s}_d)\leq z_{t;k;d},\mbox{ for all $t=1,\ldots,T$, $k=1,\ldots,K_t$ and $d=1,\ldots,D$}\}\\
 =\; & \textrm{E}_{\textbf{A}_0}\bigg\{\prod_{t=1}^{T}\textrm{E}_{\textbf{A}_{t}}\bigg(\prod_{k=1}^{K_t}\textrm{E}_{\textbf{A}_{t;k}}\bigg[\prod_{d=1}^{D}\Pr\bigg\{U_{t;k}(\mathbf{s}_{d})\leq \frac{z_{t;k;d}}{\vartheta_{t;k}(\mathbf{s}_{d})}\mid \textbf{A}_{t;k},\textbf{A}_{t},\textbf{A}_0\bigg\}\bigg]\bigg)\bigg\}\\
 =\; & \textrm{E}_{\textbf{A}_0}\bigg[\prod_{t=1}^{T}\textrm{E}_{\textbf{A}_{t}}\bigg\{\prod_{k=1}^{K_t}\textrm{E}_{\textbf{A}_{t;k}}\bigg(\prod_{d=1}^{D}\exp \bigg[-\bigg\{\frac{z_{t;k;d}}{\vartheta_{t;k}(\mathbf{s}_{d})}\bigg\}^{-\frac{1}{\alpha_{t;k}\alpha_t\alpha_0}}\bigg]\bigg)\bigg\}\bigg]\\
 =\; & \textrm{E}_{\textbf{A}_0}\bigg[\prod_{t=1}^{T}\textrm{E}_{\textbf{A}_{t}}\bigg\{\prod_{k=1}^{K_t}\textrm{E}_{\textbf{A}_{t;k}}\bigg(\exp \bigg[-\sum_{l=1}^L A_{t;k;l}\underbrace{A_{t;l}^{1/\alpha_{t;k}}A_{0;l}^{1/(\alpha_{t;k}\alpha_t)}\sum_{d=1}^{D}\bigg\{\frac{\omega_{t;k;l}(\mathbf{s}_d)}{z_{t;k;d}}\bigg\}^{1/(\alpha_{t;k}\alpha_t\alpha_0)}}_{C_{t;k;l}}\bigg]\bigg)\bigg\}\bigg]\\
 =\; &\prod_{l=1}^L\textrm{E}_{A_{0;l}}\bigg(\prod_{t=1}^{T}\textrm{E}_{A_{t;l}}\bigg[\prod_{k=1}^{K_t}\textrm{E}_{A_{t;k;l}}\bigg\{\exp\bigg(-  C_{t;k;l}A_{t;k;l}\bigg)\bigg\} \bigg]\bigg)\\
 =\; &\prod_{l=1}^L\textrm{E}_{A_{0;l}}\bigg[\prod_{t=1}^{T}\textrm{E}_{A_{t;l}}\bigg\{\prod_{k=1}^{K_t}  \exp \bigg(- C_{t;k;l}^{\alpha_{t;k}}\bigg)\bigg\}\bigg] \\
 =\; &\prod_{l=1}^L\textrm{E}_{A_{0;l}}\bigg[\prod_{t=1}^{T}\textrm{E}_{A_{t;l}}\bigg\{ \exp \bigg(-A_{t;l}\underbrace{A_{0;l}^{1/\alpha_t} \sum_{k=1}^{K_t}\bigg[\sum_{d=1}^{D}\bigg\{\frac{\omega_{t;k;l}(\mathbf{s}_d)}{z_{t;k;d}}\bigg\}^{1/(\alpha_{t;k}\alpha_t\alpha_0)}\bigg]^{\alpha_{t;k}}}_{C_{t;l}}\bigg)\bigg\}\bigg]\\
  =\; & \prod_{l=1}^L\textrm{E}_{A_{0;l}}\bigg[\prod_{t=1}^T \textrm{E}_{A_{t;l}} \bigg\{\exp\bigg(-  C_{t;l}^{\alpha_{t;l}}\bigg)\bigg\} \bigg]\\
  =\; &\prod_{l=1}^L\textrm{E}_{A_{0;l}} \bigg[\exp \bigg\{- A_{0;l}\underbrace{\sum_{t=1}^T\bigg( \sum_{k=1}^{K_t}\bigg[\sum_{d=1}^{D}\bigg\{\frac{\omega_{t;k;l}(\mathbf{s}_d)}{z_{t;k;d}}\bigg\}^{1/(\alpha_{t;k}\alpha_t\alpha_0)}\bigg]^{\alpha_{t;k}}\bigg)^{\alpha_t}}_{C_l}\bigg\}\bigg]\\
=\; &\prod_{l=1}^L \exp \left(- C_l^{\alpha_0}\right)\;=\;  \exp \bigg[-\sum_{l=1}^L  \bigg\{\sum_{t=1}^T\bigg( \sum_{k=1}^{K_t}\bigg[\sum_{d=1}^{D}\bigg\{\frac{\omega_{t;k;l}(\mathbf{s}_d)}{z_{t;k;d}}\bigg\}^{1/(\alpha_{t;k}\alpha_t\alpha_0)}\bigg]^{\alpha_{t;k}}\bigg)^{\alpha_t}\bigg\}^{\alpha_0}\bigg].
\end{align*}
Note that this also proves \eqref{eq:NRS} by setting $T=1$ and $\alpha_{1;k}\alpha_1=\alpha_k$ for all $k$.

\end{appendix}

\newpage

\begin{figure}[t!]
\begin{center}
\includegraphics[scale=0.7]{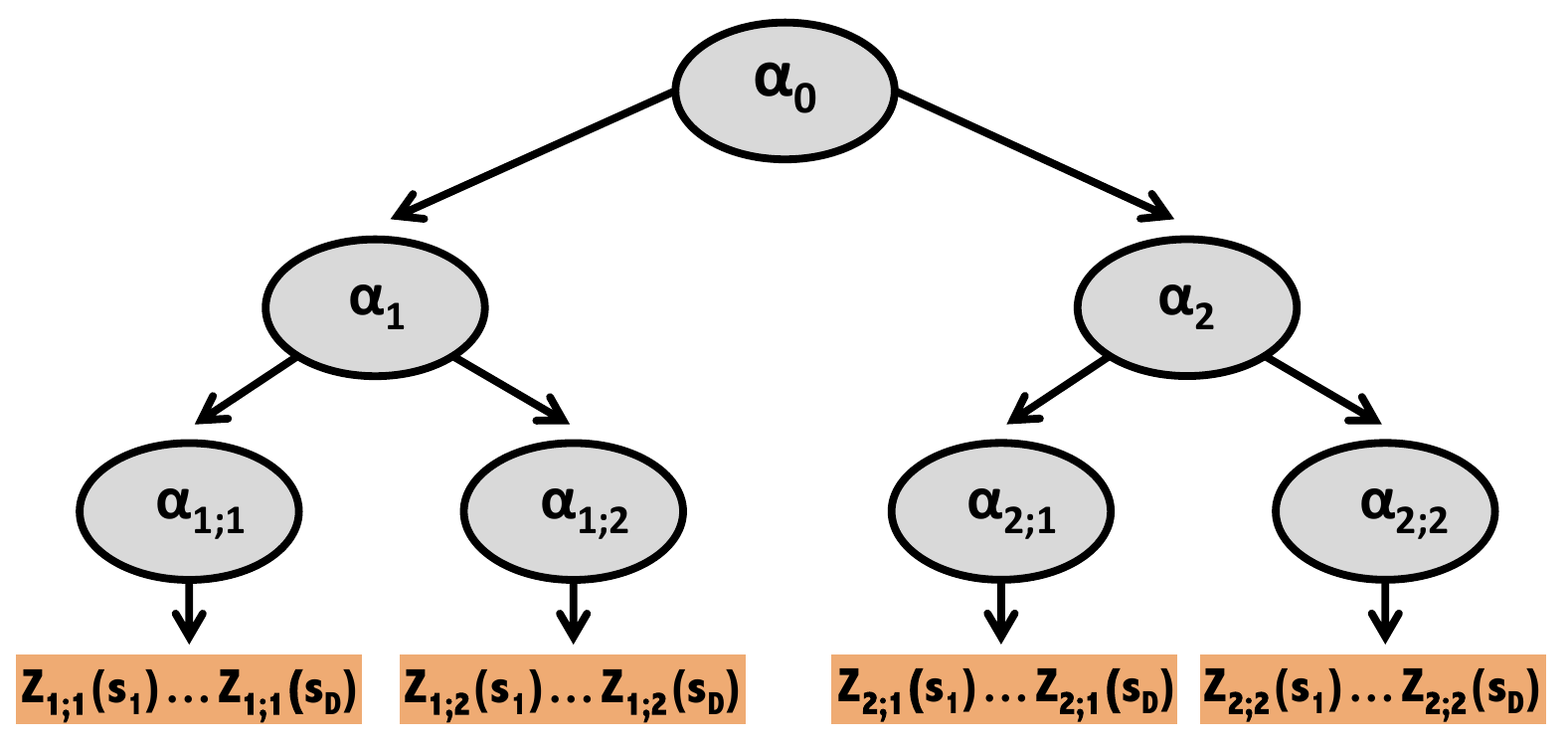}
\caption{\footnotesize{Example of a tree structure summarizing the extremal dependence of the four-dimensional process $\mathbf{Z}(\mathbf{s})=\{Z_{1;1}(\mathbf{s}),Z_{1;2}(\mathbf{s}),Z_{2;1}(\mathbf{s}),Z_{2;2}(\mathbf{s})\}^\top$. For each cluster $t=1,2$ and variable $k=1,2$, the spatial dependence of the process $Z_{t;k}(\mathbf{s})$ is summarized by the product $\alpha_{t;k}\alpha_t\alpha_0$ and the corresponding weight function $\omega_{k;t}(\mathbf{s})$. The intra-cluster cross-dependence, i.e., between processes $Z_{t;1}(\mathbf{s})$ and $Z_{t;2}(\mathbf{s})$, is summarized by the product $\alpha_t\alpha_0$. The inter-cluster cross-dependence, i.e., between the variables $Z_{t_1;k_1}(\mathbf{s}), Z_{t_2;k_2}(\mathbf{s})$, with $t_1\neq t_2$ and $k_1,k_2=1,2$, is summarized by the parameter $\alpha_0$. The number of latent $\alpha$-stable random variables involved in this model is equal to the number of upper tree nodes (excluding the terminal nodes) multiplied by the number of basis functions, $L$; here, there are $7L$ latent variables.}}\label{fig:tree_3layer}
\end{center}
\end{figure}

\begin{figure}[t!]
\begin{center}
{\centering\includegraphics[width=\linewidth]{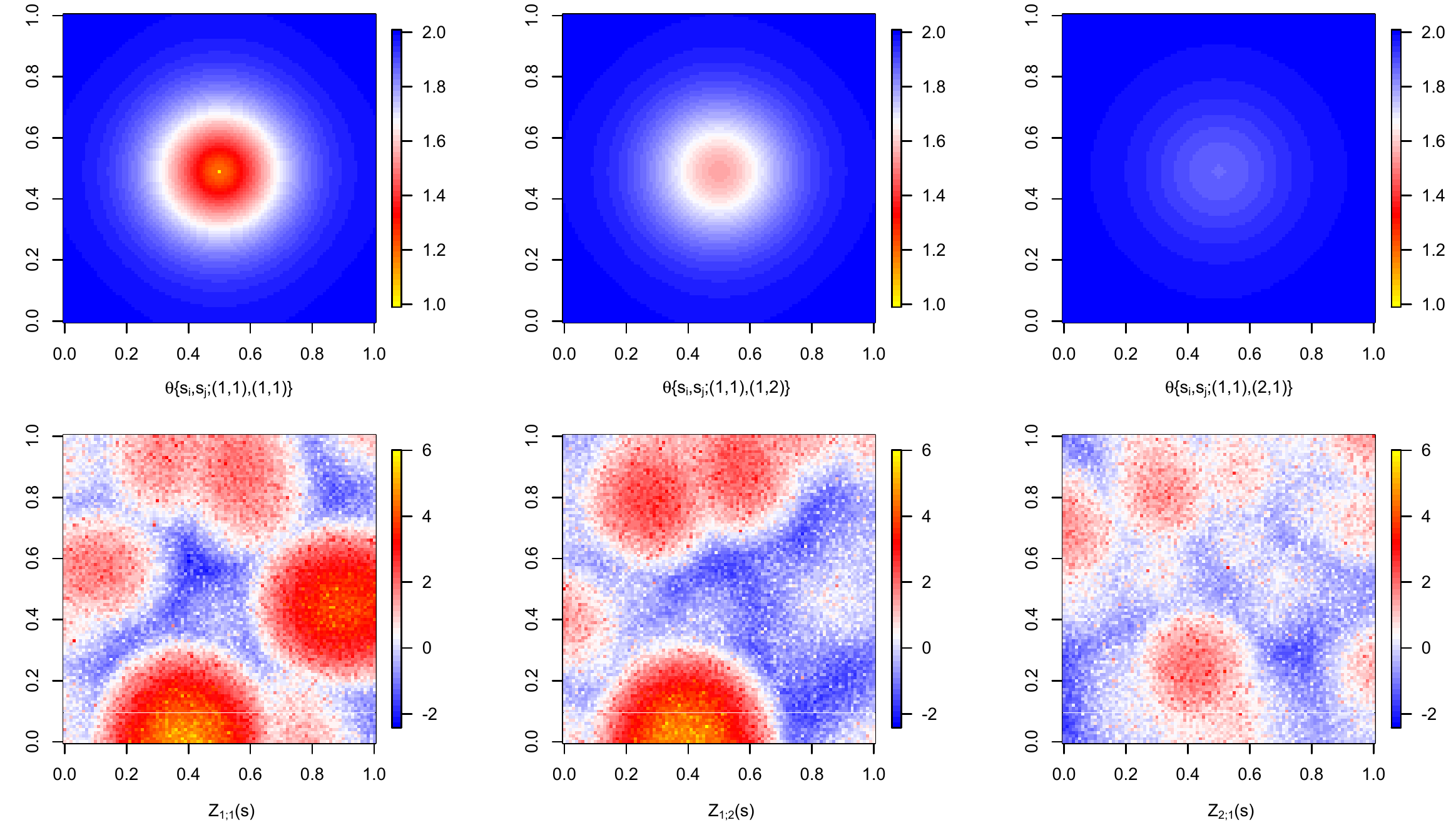}}
\end{center}
\caption{\footnotesize{\emph{Top row:} Pairwise extremal coefficient $\theta\{\mathbf{s}_i,\mathbf{s}_j;(t_1,k_1),(t_2,k_2)\}$, see \eqref{nl_extr}, for the three-layer multivariate max-stable model with exponent function \eqref{eq:NRS3} and underlying tree structure displayed in Figure~\ref{fig:tree_3layer}, for fixed reference location $\mathbf{s}_i=(0.5,0.5)^\top$ and $\mathbf{s}_j\in[0,1]^2$. Here, $\alpha_0=0.9$, $\alpha_1=\alpha_2=0.7$ and $\alpha_{1;1}= \alpha_{1;2}=\alpha_{2;1}=\alpha_{2;2}=0.4$, while the kernels are Gaussian densities as in \eqref{eq:RS} with bandwidths $\tau_{1;1}=\tau_{1;2}=\tau_{2;1} = \tau_{2;2}=0.1$, with knots taken on a $100\times100$ regular grid. The panels summarize the spatial dependence of each individual process (left), the intra-cluster cross-dependence (middle) and the inter-cluster cross-dependence (right). \emph{Bottom row:} Realizations of $Z_{1;1}(\mathbf{s})$ (left), $Z_{1;2}(\mathbf{s})$ (middle) and $Z_{2;1}(\mathbf{s})$ (right).}}\label{fig:ext}
\end{figure}

\end{document}